\documentclass[a4paper,11pt]{article}
\pdfoutput=1 

\usepackage{jcappub} 

\usepackage[T1]{fontenc} 
\usepackage{tabularx}

\title{\boldmath Models of the declining ionizing emissivity near the end of reionization}

\author[a,1]{Anson D'Aloisio,\note{Corresponding author.}}
\author[a]{Nayan A. Camalon,}
\author[b,c]{L. Y. Aaron Yung,}
\author[d]{Rachel S. Somerville,}
\author[e,f,g]{Christopher Cain,}
\author[a]{and Nakul Gangolli}

\affiliation[a]{Department of Physics and Astronomy, University of California, Riverside,
\\900 University Ave., Riverside, CA, 92521, USA}
\affiliation[b]{Space Telescope Science Institute, 3700 San Martin Drive, Baltimore, MD 21218, USA}
\affiliation[c]{Institute of Astronomy and Department of Physics, National Tsing Hua University, Hsinchu 30013, Taiwan}
\affiliation[d]{Center for Computational Astrophysics, Flatiron Institute, New York, NY 10010, USA}
\affiliation[e]{KICC - Kavli Institute for Cosmology Cambridge,
Madingley Road, CB3 0HA Cambridge, United Kingdom}
\affiliation[f]{Institute of Astronomy, University of Cambridge, Madingley Road, Cambridge CB3 0HA, UK}
\affiliation[g]{School of Earth and Space exploration, Arizona State University, Tempe, AZ 85281, USA}

\emailAdd{ansond@ucr.edu}

\abstract{Reionization is widely believed to have been driven by the rise of the ionizing radiation background from galaxy formation. Observations of high-$z$ quasars have delivered precise measurements of the Ly$\alpha$ forest mean transmission between $z=5$--6, improving constraints on the last phases of reionization considerably. Independent attempts to forward-model these measurements with radiative transfer simulations suggest that the ionizing emissivity started to decline during the latter half of reionization. The cause of this evolution remains unknown. Leading proposals include the quenching of star formation from reionization's photoheating of the IGM or a declining escape fraction ($f_{\rm esc}$).  We explore these two mechanisms using the Santa Cruz semi-analytic modeling framework.  For the IGM photoheating effect to play a significant role in producing the decline, the population-averaged escape fraction must be close to zero for galaxies with halo masses $M_{\rm halo}\gtrsim 10^{10}$ M$_\odot$. Even in this scenario, however, potentially all of the evolution comes from a combination of hierarchical growth and stellar feedback shifting the ionizing photon production toward larger $M_{\rm halo}$.  Recent radiation hydrodynamics simulations predict a rapid decline in $f_{\rm esc}$ during reionization, and we implement a prescription for this evolution from the \textsc{SPHINX} simulation project.  After an overall renormalization of $f_{\rm esc}$, we find that the \textsc{SPHINX} prediction recovers the required redshift evolution of the emissivity remarkably well.  Our results support a picture in which the declining emissivity near the end of reionization was driven mostly by $f_{\rm esc}$ evolution.}


\begin{document}
\maketitle
\flushbottom

\section{Introduction}
\label{sec:intro}

The study of quasar absorption spectra has significantly advanced our understanding of the last phases of reionization over the past decade. We now have strong evidence that reionization has been detected in the Lyman-$\alpha$ (Ly$\alpha$) and Lyman-$\beta$ forests of $z \gtrsim 6$ quasars, as well as increasingly stringent constraints on the tail end of the reionization history \cite{becker15, 2018arXiv180704229E, 2019MNRAS.485L..24K, 2019arXiv191003570N, 2019arXiv190512640K, 2021ApJ...923..223Z, 2022ApJ...932...76Z, 2022MNRAS.514...55B, 2024arXiv240508885B, 2024arXiv240512275Z}.  Concurrently, the James Webb Space Telescope (JWST) is giving us an unprecedented view of the physical conditions inside reionization-era galaxies \cite{2022A&A...665L...4S, 2023A&A...677A.115C, 2023ApJ...950...67M, 2023A&A...678A.173V}, their ionizing photon production efficiencies \cite{Endsley2023, Pahl2024, 2024MNRAS.527.6139S, 2024MNRAS.535.2998S, Papovich2025}, and their escape fractions \cite{2024ApJ...972...92J, 2024ApJ...973..111J, Papovich2025}.  A key frontier in reionization research is to better understand how high-$z$ galaxies are connected to the evolving ionization state of the intergalactic medium (IGM).  

\begin{figure}[t!]
\begin{center}
 \resizebox{7.6cm}{!}{\includegraphics[]{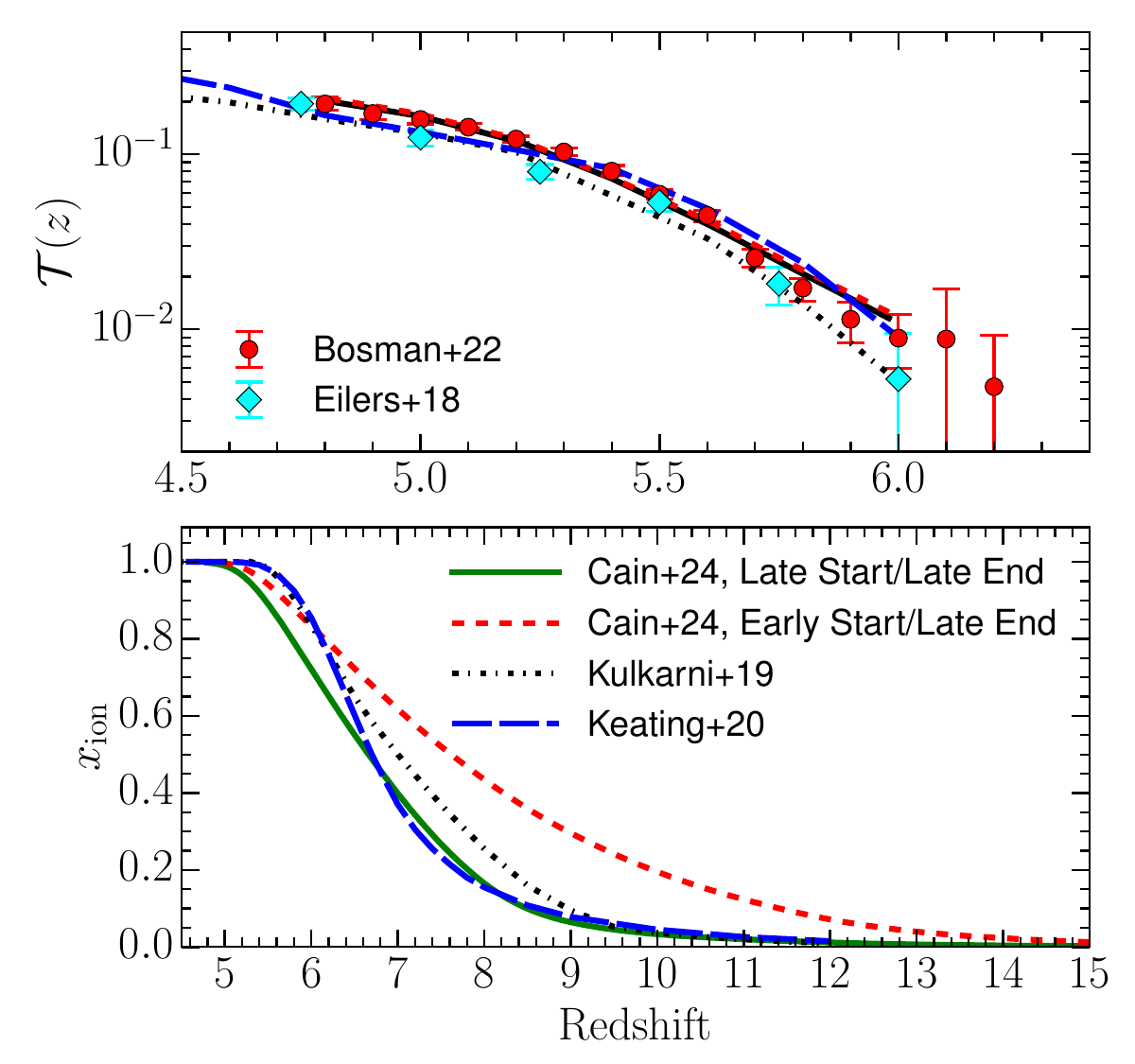}}
 \resizebox{7.6cm}{!}{\includegraphics[]{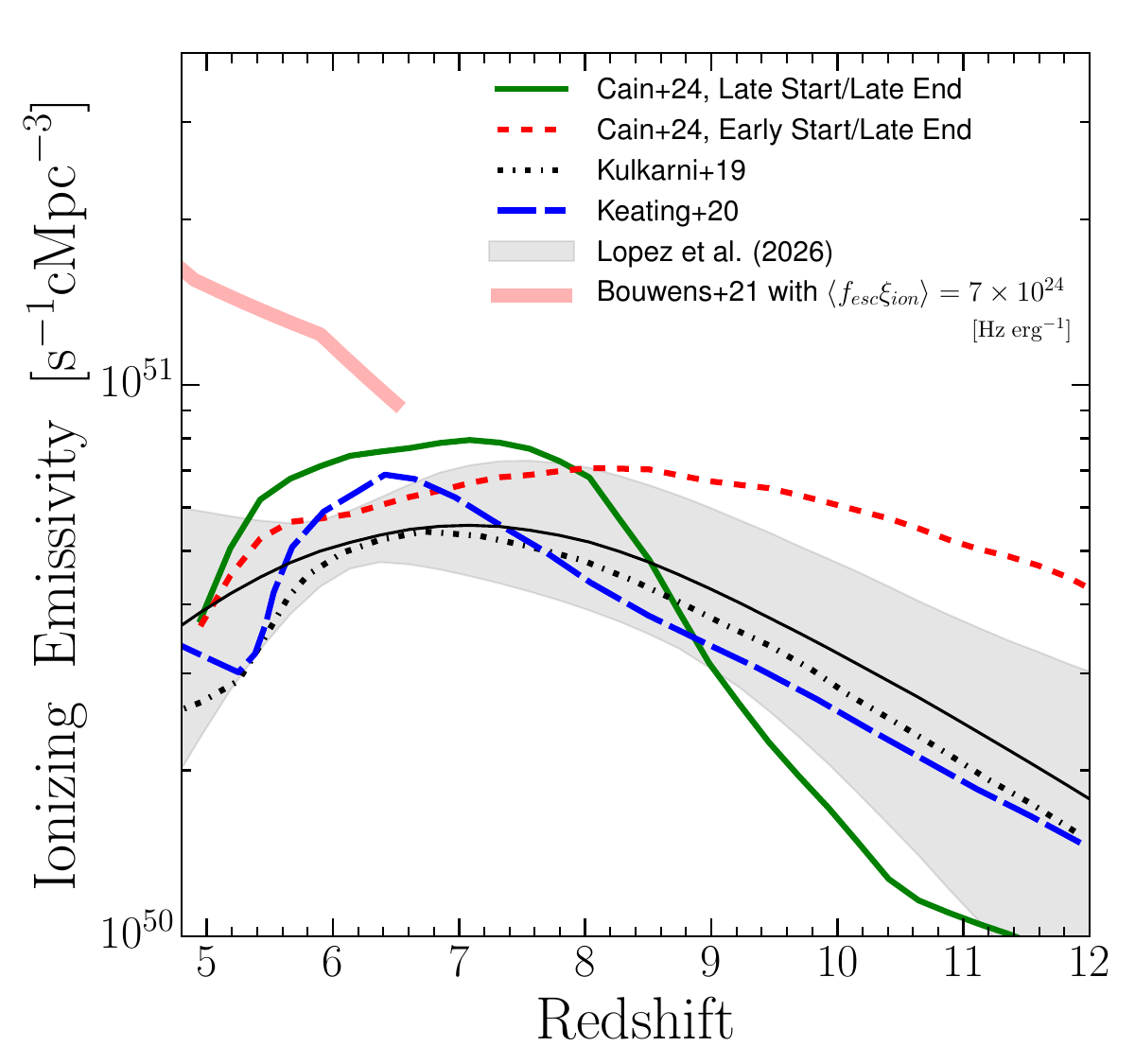}}
 \end{center}
\caption{{\bf Top Left:} The Ly$\alpha$ forest mean transmission as a function of redshift.  The data points show the measurements of Refs. \cite{2018arXiv180704229E} and \cite{2022MNRAS.514...55B}. The curves show forward-models from four RT simulations of reionization that have been calibrated to be consistent with the measurements. {\bf Bottom Left:} Reionization histories in the four simulations shown in the top-left left panel. {\bf Right:} The curves show ionizing emissivities in the four calibrated reionization simulations from the left panels. The thin black curve and shaded gray region show the median and 95 \% credible levels from the joint forest-CMB analysis from \cite{2026arXiv261028765L}. Their results adopt a fixed model for the opacity of the IGM to ionizing photons, as well as a fixed cosmology.  The key point is that the emissivities rise and decline near the end of reionization, an evolution that contrasts with the measured non-ionizing UV luminosity density, $\rho_{\rm UV}$, which increases monotonically with time during the same epoch. The thick pink curve-segment shows, for illustrative purposes, the emissivity obtained from the measured $\rho_{\rm UV}$ of \cite{2021AJ....162...47B}, assuming a constant $\langle f_{\rm esc} \xi_{\rm ion} \rangle = 7 \times 10^{24}$ Hz erg$^{-1}$. Understanding the apparent rise and decline of the ionizing source population is the central focus of this paper.  }
\label{fig1} 
\end{figure}

The most basic Ly$\alpha$ forest statistic is the mean transmission fraction, $\bar{\mathcal{T}}(z)$, defined to be the fraction of the quasar continuum that is not absorbed by hydrogen in the IGM, averaged over the forest at redshift $z$. The data points in the top-left panel of Figure \ref{fig1} show two independent measurements of $\bar{\mathcal{T}}(z)$ by Refs. \cite{2018arXiv180704229E} and \cite{2022MNRAS.514...55B}. These measurements provide important ``boundary conditions'' for reionization models.  A realistic model of reionization must conclude in a way such that $\bar{\mathcal{T}}(z)$ is consistent with the observed values. Indeed, many reionization simulations nowadays are calibrated to achieve such agreement. In some simulations, this involves adjusting the redshift evolution of the ionizing emissivity, or rate of ionizing photons per unit volume emitted into the IGM, until the simulated $\bar{\mathcal{T}}(z)$ agrees with measurements. Other more physically detailed simulations, e.g. Refs. \cite{2020MNRAS.496.4087O, 2022MNRAS.511.4005K, 2014ApJ...793...29G}, model galaxy formation in a bottom-up way with prescriptions for star formation and feedback processes. In this case, the star formation and feedback prescriptions are calibrated also to other observables, usually of the galaxy population, such that the simulated galaxies resemble as closely as possible the observed galaxy population. 

Measurements of $\bar{\mathcal{T}}(z)$ have played a central role in developing our understanding of reionization's last phases.   The bottom-left panel of Figure \ref{fig1} shows the volume-averaged ionized fraction, $x_{\rm ion}$, in four reionization models from Refs. \cite{2019MNRAS.485L..24K, 2020MNRAS.491.1736K, 2024arXiv240902989C} that have been calibrated in the first manner described above.  The curves in the top-left panel show the agreement between the simulated and observed $\bar{\mathcal{T}}(z)$.  Note that all models have neutral fractions of $1 - x_{\rm ion} \sim 15-30\%$ at $z=6$, with reionization concluding around $z=5-5.4$. Indeed, the first attempts at calibrating cosmological reionization simulations with $\bar{\mathcal{T}}(z)$ led to the realization that reionization must have extended below $z=6$.  This naturally accounts for large fluctuations in the effective optical depth, $\tau_{\rm eff}$ \cite{2019MNRAS.485L..24K}.  Within a few years, a consensus formed around ``late-ending reionization'' models, as it was shown that they are broadly consistent with all existing quasar absorption spectra statistics \cite{2019MNRAS.485L..24K, 2019arXiv190512640K, 2019arXiv191003570N, 2021MNRAS.508.1853B, 2021ApJ...923..223Z, 2022ApJ...932...76Z, 2022MNRAS.514...55B, 2023ApJ...955..115Z, 2024A&A...688L..26S, 2024arXiv240508885B, 2024arXiv240512275Z}.  But the success of late reionization models has led to another puzzle about the sources.  The right panel of Figure \ref{fig1} shows the calibrated ionizing emissivities of the same four simulations in the left panels.\footnote{It is worth noting that two independent research groups with different radiative transfer methodologies are represented in this figure. The Cain+24 simulations used an adaptive ray tracing code, while the Kulkarni+19 and Keating+20 \cite{2019arXiv190512640K} simulations used \textsc{ATON}, an M1-closure moment-based code \cite{2008MNRAS.387..295A}. The declining emissivity is also seen in \textsc{CoDa~III}, {an example of a ``bottom-up'' radiative hydrodynamics simulation that also attempts to model galaxy formation} (P. Ovirk, personal communication, June 2024).  To our knowledge, the only RT simulation in the literature that agrees (broadly) with $\bar{\mathcal{T}}(z)$ measurements and does not exhibit an emissivity decline is \textsc{THESAN} \cite{2022MNRAS.511.4005K, 2022MNRAS.512.4909G}. However, Ref. \cite{2024MNRAS.531.1951C} showed that this likely owes to \textsc{THESAN}'s use of the reduced speed of light approximation, $c_{\rm sim} = 0.2c$. A reduced speed of light artificially suppresses the growth of the ionizing background and the propagation of ionization fronts near the end of reionization \cite{2019A&A...622A.142D, 2019A&A...626A..77O}, alleviating the need for the emissivity to decline.} The gray shading and thin curve show model-dependent constraints on the emissivity from a joint Bayesian MCMC analysis with Ly$\alpha$ forest and {\it Planck} TTTEEE likelihoods within the semi-analytic modeling framework of \cite{2026arXiv260619449C} \cite{2026arXiv261028765L}. {\bf Consistency with the measured $\bar{\mathcal{T}}(z)$ broadly requires there to be a turnover in the emissivity at $z\simeq 6-8$.} 

Ref. \cite{2024MNRAS.531.1951C} explored several potential ways that radiative transfer simulations could fail to model the relationship between the emissivity and $\bar{\mathcal{T}}(z)$, resulting in a spurious decline of the former.  These included missing opacity from circumgalactic gas around massive galaxies, small-scale IGM clumpiness, and incorrect assumptions about the spectrum of the sources. They found that, while a combination of these effects could plausibly flatten the rate of decline, it is generally difficult to construct a model with monotonically rising emissivity that agrees with measurements of $\bar{\mathcal{T}}(z)$.   Indeed, the evolution of the emissivity does not appear to trace the measured non-ionizing UV luminosity density of the Universe, $\rho_{\rm UV}$, which is observed to rise steeply during the same epoch \cite{2021AJ....162...47B}.\footnote{{We note that current measurements of $\rho_{\rm UV}$ at $z>5$ extend down to $M_{\rm UV}\sim -17$ and may not encompass the galaxy population that produces many of the ionizing photons \cite[e.g.][]{2020MNRAS.494.1002Y, 2020MNRAS.tmp.1918Y}.}}  The galaxy population's escaped ionizing emissivity is often written as

\begin{equation}
\dot{n}^{\rm esc}_{\rm ion} = \langle f_{\rm esc} \xi_{\rm ion} \rangle \cdot \rho_{\rm UV},
\label{eq1}
\end{equation}
where $f_{\rm esc}$ is the escape fraction, and $\xi_{\rm ion}$ is the ionizing efficiency, and $\langle \dots \rangle$ denotes a UV-luminosity-weighted average over the population. The thick pink curve segment in the right-panel of Figure \ref{fig1} shows the ionizing emissivity derived from the measurements of $\rho_{\rm UV}$ by \cite{2021AJ....162...47B}, adopting for illustrative purposes a constant $\langle f_{\rm esc} \xi_{\rm ion} \rangle = 7 \times 10^{24}$ Hz erg$^{-1}$ and integrating the rest-frame ultraviolet (UV) luminosity function down to $M_{\rm UV} = -16$. \footnote{This requires modestly extrapolating the faint-end of their UV luminosity function measurement. However, our key point that the UV luminosity density of the Universe was rising with time at this epoch does not depend on the details of the extrapolation.} 

{\it Why does the ionizing emissivity appear to decline near the end of reionization, and what is the decline telling us about the sources of the ionizing radiation background at $z\sim 6$?} The current paper aims to address these two broad questions.  Two distinct explanations have been proposed to date.   Ref. \cite{2021MNRAS.507.6108O} proposed that the evolution is caused by reionization's photo-heating of the IGM, which is believed to suppress star formation in low-mass halos \cite[e.g.][]{1992MNRAS.256P..43E, shapiro94, gnedin00, noh14, 2020MNRAS.494.2200K}.  In what follows, we will refer to this effect as ``reionization feedback.'' In this scenario, reionization is driven by low-mass galaxies which suffer a decline in their ionizing photon output because of the progression of reionization itself. Alternatively, Ref. \cite{2022MNRAS.515.2386R} proposed that the declining emissivity results from a rapid evolution in the escape fractions of galaxies.  This proposal is based on results from the \textsc{SPHINX} simulations \cite{2018MNRAS.479..994R}, which suggest that escape fractions trace the evolution of the specific star formation rates of the galaxy population. Ionizing photons are thought to escape through low-column-density channels in the interstellar medium (ISM) created by strong stellar feedback events \cite{2014ApJ...788..121K,2015MNRAS.453..960M,2017MNRAS.470..224T,2022MNRAS.515.2386R,2023MNRAS.518..270K}. In the proposal of \cite{2022MNRAS.515.2386R}, decreasing specific star formation rates indicate a declining efficacy in creating such channels, hence a declining escape fraction.  Here we will explore both proposals to assess their potential roles in generating the non-monotonic emissivity evolution illustrated in Figure \ref{fig1}.   

The remainder of the paper is organized as follows.  In \S \ref{sec:methods}, we describe our modeling methods. In \S \ref{sec:results}, we present our main results. In \S \ref{sec:conclusion} we offer concluding remarks.  We adopt a standard $\Lambda$CDM cosmology with parameters $\Omega_m = 0.31$, $\Omega_\Lambda = 0.69$, $\Omega_b = 0.048$, $\sigma_8 = 0.82$, and $n_s = 0.96$, consistent with the latest measurements \cite{2020A&A...641A...6P}.  Unless otherwise noted, all distances are reported in comoving units.

\section{Semi-Analytic Galaxy Formation Model}
\label{sec:methods}

\subsection{Santa Cruz SAM}

We model the galaxy population with the Santa Cruz semi-analytic model (SAM), which uses the mass accretion histories of halos with empirically calibrated prescriptions to model star formation and feedback processes. The Santa Cruz SAM has been described extensively elsewhere \cite{1999MNRAS.310.1087S, Somerville2008, Somerville2012, Popping2014, 2014MNRAS.444..942P, 2015MNRAS.453.4337S, 2019MNRAS.483.2983Y, Yung2019b}. This work builds in particular on Refs. \cite{2020MNRAS.494.1002Y, 2020MNRAS.tmp.1918Y}. Following \cite{2024MNRAS.530.4868Y, 2026MNRAS.553g1913Y}, we apply the SAM to halo merger histories constructed from the Very Small MultiDark Planck simulation (\textsc{VSMDPL}) \cite{2016MNRAS.457.4340K}. \textsc{VSMDPL} has a box size of $L_{\rm box} = 160~h^{-1}$ cMpc and $N_p=3840^3$ dark matter particles. Halos identified with the \textsc{Rockstar} phase-space halo finder \cite{2013ApJ...762..109B} were obtained from the \textsc{CosmoSim Database}\footnote{\url{https://www.cosmosim.org/metadata/vsmdpl/}}, and merger trees were constructed using \textsc{Consistent Trees} \cite{2013ApJ...763...18B}.  The minimum halo mass in our models is $5\times10^8$ M$_\odot$.  Previous studies using the VSMDPL simulation have found approximate convergence in the halo mass function above this threshold \cite{2022MNRAS.509..595L, 2024MNRAS.530.4868Y, 2020arXiv200408401H}.   For ``normal'' star formation, we use the Kennicutt-Schmidt prescription described in \cite{1999MNRAS.310.1087S, Somerville2008}, in which the star formation rate surface density scales with the surface density of cold neutral gas above a threshold $\Sigma_{\rm crit} = 6$ M$_\odot$ pc$^{-2}$.  Merger-driven starbursts are modeled by converting a fraction of the available cold gas into stars with an efficiency that depends on the mass ratios and gas content of the merging galaxies.  We also include the ``stars+gas'' disk-instability starburst mode described in Ref. \cite{2014MNRAS.444..942P}.  

The parameters of the SAM are calibrated to reproduce properties of the local galaxy population \cite{2019MNRAS.483.2983Y, 2026MNRAS.553g1913Y}, including the stellar mass function \cite{2012MNRAS.421..621B, 2013MNRAS.436..697B, 2013ApJ...767...50M}, mass--metallicity relation \cite{2005MNRAS.362...41G, 2011ApJ...727...78K}, stellar-to-halo mass relation \cite{2017MNRAS.470..651R}, cold gas fractions \cite{2018RMxAA..54..443C, 2018MNRAS.476..875C}, and black hole mass--bulge mass relation \cite{2013ARA&A..51..511K, 2013ApJ...764..184M}. The SAM models stellar feedback with a power-law parametrization for the mass outflow rate, $\dot{m} = \epsilon_{\rm SN} (V_c/200~\mathrm{km~s}^{-1})^{-\alpha_{\rm rh}}~\dot{m}_*$, where $\dot{m}_*$ is the star formation rate of a galaxy, $V_c$ is its circular velocity, and $\epsilon_{\rm SN}$ and $\alpha_{\rm rh}$ are free parameters \cite{2019MNRAS.483.2983Y}.   Following previous implementations of the model, we adopt $\epsilon_{\rm SN} = 1.7$ and $\alpha_{\rm rh} = 3$, and assume that gas is ejected if $V_c < 110$ km s$^{-1}$. {These parameters were adopted to match the observed abundance of low-mass galaxies and their stellar mass fractions \cite{2019MNRAS.483.2983Y}. They have also been shown to broadly reproduce observed stellar masses and star formation rates at $4\lesssim z\lesssim8$ \cite{Yung2019b}, and rest-frame UV luminosity functions at $4\lesssim z\lesssim10$ \cite{2019MNRAS.483.2983Y}.}  We also include both radiative- and radio-mode feedback from active galactic nuclei (AGN) following Ref. \cite{Somerville2008}, although we find that it has little effect at the redshifts of interest for this work (see Appendix \ref{app:feedback}).

\subsection{Reionization Feedback Models}

\subsubsection{Noh \& McQuinn (2014)}

One of our primary focuses is the role of reionization feedback in quenching the gas accretion of star forming halos.  We have modified the Santa Cruz SAM to implement a more detailed model for reionization feedback using the prescription of Ref. \cite{noh14}. They used high-resolution hydrodynamic simulations to study the physical conditions that determine whether photoionization feedback halts a halo's gas accretion.  They found that a gas parcel is likely prevented from accreting if its Jeans mass becomes significantly larger than the halo virial mass at any point along its history in temperature-density phase space.  From this insight, they developed a merger-tree-based model that can be applied to SAMs, which we implement into the gas accretion rates computed for each halo individually.  

{We implement the reionization history externally to the SAM, i.e. it is not computed self-consistently from the ionizing emissivity of the galaxy population.}  We assign reionization redshifts, $z^{\rm root}_{\rm re}$, to each root halo at $z=0$ in the merger tree catalog. The values are drawn randomly with uniform probability between $z=5.5$ and $12$.  The progenitors within a tree are then randomly assigned their own $z_{\rm re}$ within an interval $\Delta z = 1$ centered on $z^{\rm root}_{\rm re}$. Overall this yields an effective reionization history (cumulative distribution of $z_{\rm re}$) that is approximately linear in $z$ between $z\sim 5.5$ and $12$ -- for our purposes, a reasonable approximation to the late reionization models favored by observations \cite{2025PASA...42...49Q}. To justify this procedure, consider that the time-scale for an ionization front (I-front) traveling at velocity $v_{\rm IF}$ to cross the Lagrangian radius of a halo with mass $M_{\rm halo}$ is

\begin{equation}
\Delta t_{\rm cross} \simeq 245~\mathrm{Myr}~\left(\frac{M_{\rm halo}}{10^{12}~\mathrm{M}_{\odot}} \right)^{1/3} \left( \frac{v_{\rm IF}}{10^3~\mathrm{km}~\mathrm{s}^{-1}}\right)^{-1}\left( \frac{1+z}{8} \right)^{-1}     
\end{equation}
I-front speeds are expected to be in the range $1,000-10,000$ km/s, where the lower and upper values are representative of the beginning and end of reionization, respectively \cite{daloisio18}.  At $z=7$, the crossing timescale translates to a redshift interval of $\Delta z \sim 0.15-1.4$ a for present-day $M_{\rm halo}\sim 10^{12}$ M$_\odot$, with the longer intervals in this range being representative of cosmic overdensities.  This estimate is broadly consistent with simulations, which suggest that many satellites of present-day galaxies were reionized at similar redshifts \cite[e.g.][]{2018MNRAS.477..867D}.  We note that the procedure of randomly assigning $z^{\rm root}_{\rm re}$ neglects its correlation with $M_{\rm halo}$ from halo clustering, {as well as its dependence on assembly history at fixed $M_{\rm halo}$}.  In Appendix \ref{app:zre}, we explore the correlation between $z^{\rm root}_{\rm re}$ and $M_{\rm halo}$ using a previously-run radiative transfer simulation of reionization from Ref. \cite{2024arXiv240902989C}.  We find that a halo's reionization redshift depends only mildly on its mass for $M_{\rm halo}\lesssim 10^{10}$ M$_\odot$. As we will see, reionization feedback has little effect on halos above this mass scale, justifying our approximation to neglect the $M_{\rm halo}$-dependence of $z^{\rm root}_{\rm re}$.\footnote{As shown in Figure \ref{fig:Reo_timesappendix}, the majority of low-mass halos that are susceptible to reionization feedback are reionized later than average.  Drawing $z_{\rm re}$ from the $M_{\rm halo}\lesssim10^{10}$ M$_\odot$ probability distribution shown there (instead of the uniform $z=5.5$--12 probability adopted here) would diminish the effect of reionization feedback because of the lower values of $z_{\rm re}$. In what follows, we use our fiducial model to argue that reionization feedback likely has a subdominant effect on the emissivity evolution. In this sense, our choice of reionization history is  a conservative one because it likely over-estimates the effect somewhat.   } 

Following Ref. \cite{noh14}, we model the temperature-density trajectories of collapsing gas around each halo assuming spherical collapse.  The physical outline of the model is as follows. In the absence of reionization, the gas expands until breaking away from the Hubble flow at $z_{\rm ta}$ (the turnaround redshift).  After this, it collapses adiabatically until reaching temperatures/densities where the radiative cooling timescale is shorter than the dynamical time.  Following Ref. \cite{noh14}, we compute this characteristic temperature, $T_{\rm cool}(n_{\rm H})$, as a function of the hydrogen number density, by setting the net cooling time equal to the dynamical time.\footnote{Specifically, we include collisional excitation, recombination, free-free, and collisional ionization cooling for gas of primordial composition, and photo-heating from hydrogen and singly ionized helium. Following \cite{noh14}, we adopt a flat spectrum between 1 and 4 Ry and a photoionization rate of $\Gamma_{\rm HI} = 10^{-13}$ s$^{-1}$, representative of values in simulations of reionization and in Ly$\alpha$ forest measurements \cite[e.g.][]{Gaikwad2023, 2021ApJ...917L..37C}.  We neglect Compton cooling, which affects $T_{\rm cool}(n_{\rm H})$ only at low densities, $n_{\rm H}\simeq 5\times10^{-4}$ cm$^{-3}$, which do not come into play in our calculations of the suppression mass-scale during reionization.}  When the gas reaches these temperatures/densities, it cools very efficiently and collapses to high densities inside the halo. 

The key insight is that the gas' ability to accrete depends on the local timing of reionization along its temperature-density trajectory, as well as the mass of the halo.  Using hydrodynamic simulations, Ref. \cite{noh14} found that accretion is likely halted if the instantaneous Jeans mass becomes significantly larger than the halo virial mass at any point along the trajectory.  Reionization impulsively heats the gas to $T\sim 20,000$ K at a redshift $z_{\rm re}$. If $z_{\rm re} > z_{\rm ta}$, the gas cools to a temperature floor of T$\sim 10,000$ K before turning around. Hence, for these cases, the model of \cite{noh14} assumes that the gas is at temperature $T=10,000$ K until $z_{\rm ta}$.  The maximum Jeans mass along the trajectory, $M_{\rm J}^{\rm max}$, occurs when the adiabatic collapse intersects $T_{\rm cool}(n_{\rm H})$. We solve for $M_{\rm J}^{\rm max}$ numerically and the halo is allowed to accrete new hot gas if $M_{\rm halo} > M_{\rm J}^{\rm max}/4$, where the factor of $1/4$ is a calibration factor obtained by comparing the model to halo gas fractions in simulations \cite{noh14}.  If, on the other hand, $z_{\rm re} < z_{\rm ta}$, the gas is allowed to accrete if $M_{\rm halo} > M_{\rm J}(n_{\rm H}^{\rm re}, T_{\rm cool}(n_{\rm H}^{\rm re}))/4$, where $n_{\rm H}^{\rm re}$ is hydrogen number density evaluated at $z_{\rm re}$. Ref. \cite{noh14} found that this is a reasonable approximation to cases where the gas is reionized late during its collapse trajectory. They also incorporated a threshold below which halos cannot retain gas owing to photoevaporation, which they nominally set to $10^{8}$ M$_\odot$. We do not implement such a threshold here since our simulations do not resolve halos below this mass scale. Note that star formation generally is not quenched instantaneously in this model; even if accretion is halted, star formation will continue until the galaxy has depleted its remaining cold interstellar medium (ISM) gas reservoir.  Moreover, a galaxy can later acquire additional cold gas through mergers with its satellite galaxies. 

\begin{figure}
    \centering
    \includegraphics[width=0.7\columnwidth]{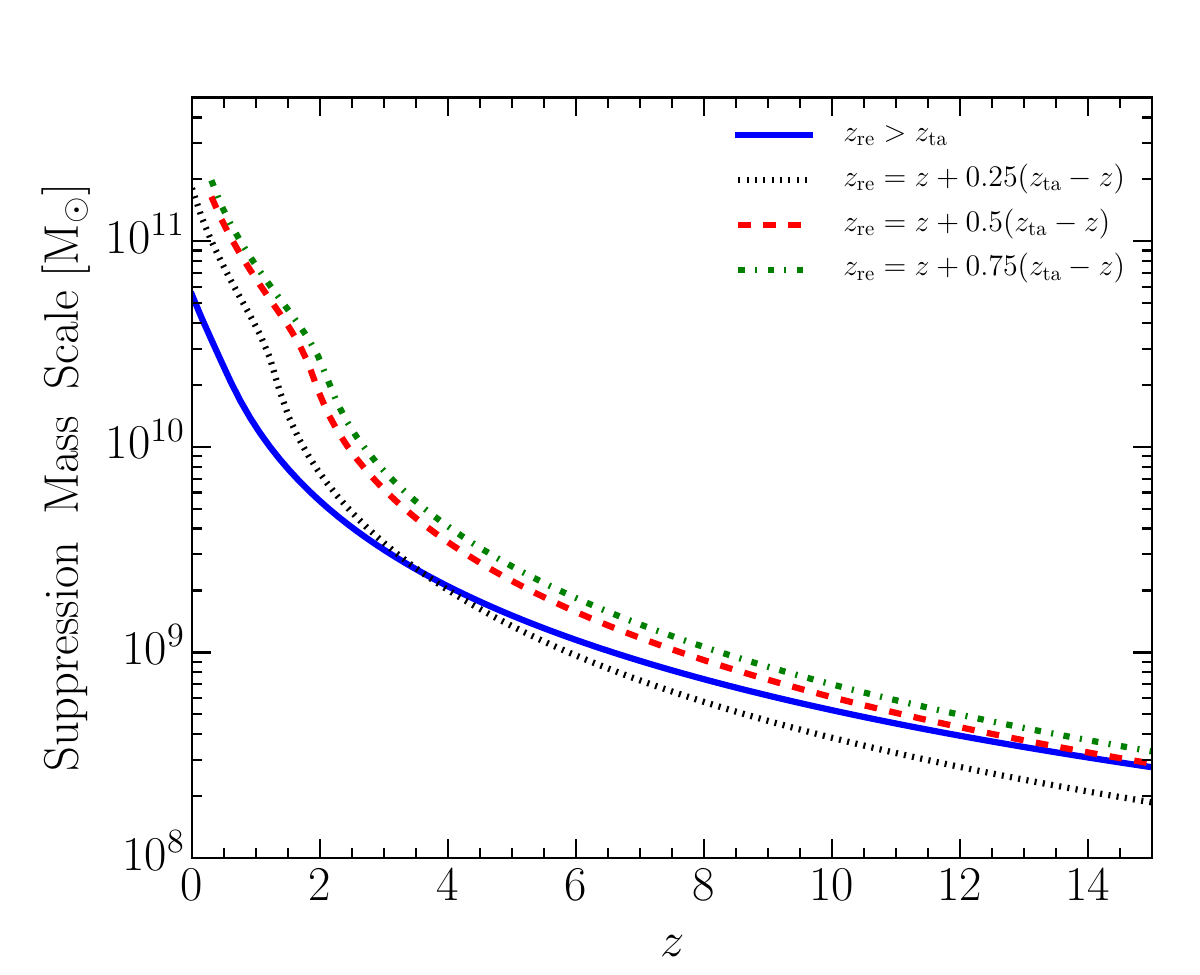}
    \caption{Characteristic halo mass scale below which gas accretion is suppressed in the \textsc{Noh \& McQuinn} reionization feedback model, as function of collapse redshift $z$. The model is based on the Jeans mass along $T$--$n_{\rm H}$ trajectories for spherically collapsing gas. The local redshift of reionization for a halo is denoted by $z_{\rm re}$ and the spherical collapse turnaround redshift is $z_{\rm ta}$. For collapsing gas that is reionized before turnaround ($z_{\rm re} > z_{\rm ta}$; blue/solid), this characteristic halo mass is one-fourth of the maximum Jeans mass along the collapse trajectory, where the one-fourth is a calibration factor obtained by comparison of the model to the results of hydrodynamic simulations in Ref. \cite{noh14}. The remaining curves show examples cases for which reionization happens between turnaround and collapse, $z < z_{\rm re} < z_{\rm ta}$ (see main text for details). During reionization, the suppression scale below which reionization feedback suppresses star formation is $M_{\rm halo} \simeq 5\times10^8 - 2\times 10^9$ M$_\odot$.  } 
    \label{fig:Mchar}
\end{figure}

Figure \ref{fig:Mchar} shows the halo mass-scale below which accretion is shut off in the \textsc{Noh \& McQuinn} model, as a function of collapse redshift $z$.  For collapsing gas that is reionized before the turnaround redshift ($z_{\rm re} > z_{\rm ta}$, blue/solid curve), the suppression scale $M^{\rm max}_{\rm J}/4$ is independent of $z_{\rm re}$ in the model, so it depends only on the collapse redshift.  The three other curves correspond to situations where reionization happens after turnaround.  In all cases, the suppression scale is within the range $M_{\rm halo} \simeq 5\times10^8 - 2\times 10^9$ M$_\odot$ during reionization.

\subsubsection{Okamoto et al. (2008) $M_c$ Model}

In previous applications, the Santa Cruz SAM applied a baryon-retention prescription, based on Refs. \cite{2000ApJ...542..535G, 2004ApJ...609..482K, 2008MNRAS.390..920O}, in which the fraction of baryons that collapse into a halo of mass $M_{\rm halo}$ at redshift $z$ is

\begin{equation}
f_b(M_{\rm halo},z) = \langle f_b \rangle \left\{ 1+(2^{\alpha/3} -1 )\left( \frac{M_{\rm halo}}{M_{\rm c}(z)}\right)^{-\alpha}\right\}^{-3/\alpha},
\end{equation}
where $\langle f_b \rangle$ is the cosmic mean baryon fraction, and $M_{\rm c}(z)$ is a characteristic mass scale. For the purpose of comparison, we have performed runs with this model as well. Following previous Santa Cruz SAM implementations, we adopt $\alpha = 2$ and assume that all halos are reionized at $z_{\rm re}=8$.  We apply the generalized analytic approximation for $M_{\rm c}(z)$ from Ref. \cite{2022MNRAS.514.2667K}.  We use their equation (9) with sharpness parameter $\gamma=15$.   This model, which we term the \textsc{Okamoto $M_c$} model, serves as a useful comparison of previous implementations to our more detailed treatment of reionization feedback.  With $\alpha =2$, $M_{\rm c}(z)$ corresponds to the mass at which halos accrete one-half of the cosmic mean baryon fraction.  At $z=6$, $7$, and $8$, $M_{\rm c}$ takes on values of $1.9\times10^8$, $8.6\times10^7$, and $9.3\times10^6$ M$_\odot$, respectively. Based on these numbers in comparison to Figure \ref{fig:Mchar}, we can anticipate that this model will yield substantially weaker reionization feedback effects compared to our fiducial \textsc{Noh \& McQuinn} model.  

\subsection{Stellar Populations and Ionizing Emissivities}
\label{sec:SPS}
 
{The Santa Cruz SAM framework computes the ionizing emissivity self-consistently, including the effects of stellar age and metallicity, i.e. we do not adopt the common practice of assuming a fixed value for $\xi_{\rm ion}$.\footnote{{See \cite{2020MNRAS.494.1002Y} for a discussion of $\xi_{\rm ion}$ in the SAM.}}} We use the Binary Population and Spectral Synthesis (BPASS version 2.2.1) model to construct spectral properties for the SAM galaxies \cite{2017PASA...34...58E}.   We adopt a Chabrier initial mass function \cite{2003PASP..115..763C} in the mass range $0.1 \leq M \leq 300~\mathrm{M}_\odot$, with a low-mass turnover at $M\sim1~\mathrm{M}_\odot$ and a  power-law slope of $-2.3$ in the large-mass regime. The two spectral properties of interest for this work are the intrinsic ionizing emissivity and rest-frame ultraviolet (UV) luminosity.  We pre-compute these quantities for stellar populations on a grid of absolute metallicity bins between $Z = 10^{-5}-0.04$, and logarithmically spaced age bins with $\Delta \log_{10}t/[\mathrm{Myr}] = 0.1$.  For UV luminosities, we apply a top-hat filter of width $400$\AA\ centered at 1600 \AA.  For each galaxy the SAM outputs star formation histories (SFHs) in 12 bins of stellar metallicity and cosmic time bins of width $\Delta t = 10$ Myr. We first rebin the SFHs onto the BPASS stellar-age grid for each metallicity bin, conserving the amount of stellar mass formed in each time interval.  We then convolve the re-binned SAM star-formation rate as a function of stellar age with the BPASS quantities per unit stellar mass.  Finally, the results are summed over the metallicity bins to obtain the intrinsic ionizing emissivity and UV luminosity of each galaxy. {For the  latter, we apply dust attenuation using the metallicity, the mass and radius of the cold gas disk, and a randomly drawn inclination angle with a slab model in which the stars are embedded in a slab of dust (see \S 2.5 of \cite{2019MNRAS.483.2983Y}). This approach has been shown to yield rest-frame UV continuum slopes consistent with observationally measured values using JWST \cite{2019MNRAS.483.2983Y, 2023MNRAS.520...14C}.}

\section{Results}

\label{sec:results}

\begin{figure}
    \centering
    \includegraphics[width=\columnwidth]{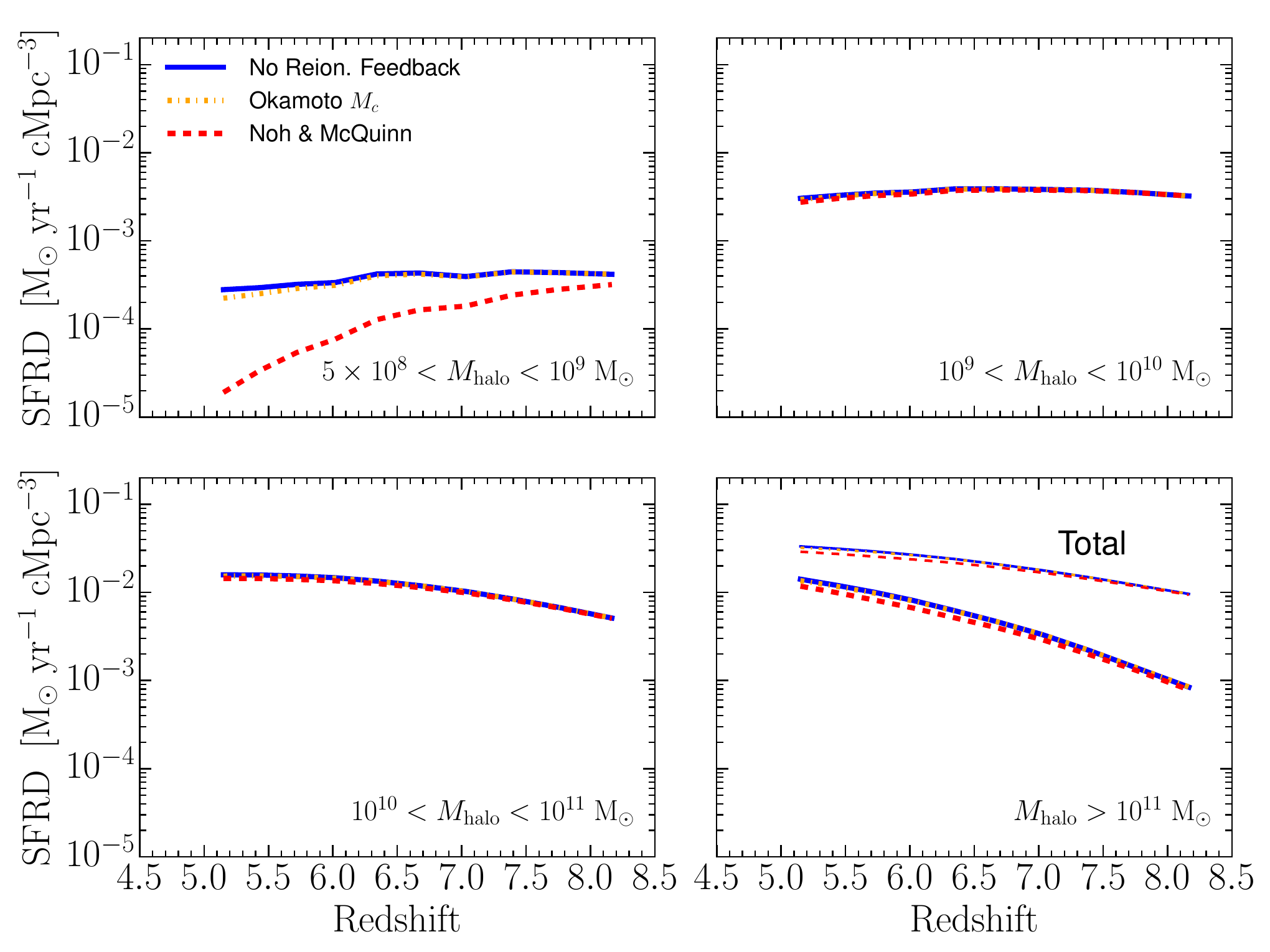}
    \caption{The star formation rate density of halos in our SAM runs for the reionization feedback models considered here, binned by halo mass. We consider three scenarios: (1) a case in which reionization photoheating has no effect on gas accretion (\textsc{No Reion. Feedback}); (2) the baryon-retention prescription based on the characteristic mass of Ref. \cite{2008MNRAS.390..920O}, $M_c$, which corresponds to the mass at which halos accrete half of the cosmic mean baryon fraction (\textsc{Okamoto $M_c$}); (3) our fiducial \textsc{Noh \& McQuinn} model.  For reference, the thin curves in the top of the bottom-right panel show the total SFRD. Reionization feedback has a significant effect on the SFRD only in the lowest mass bin, for halos in the range $5\times 10^8 < M_{\rm halo} < 10^9$ M$_\odot$.   }
    \label{fig:sfrd}
\end{figure}

\begin{figure}
    \centering
    \includegraphics[width=\columnwidth]{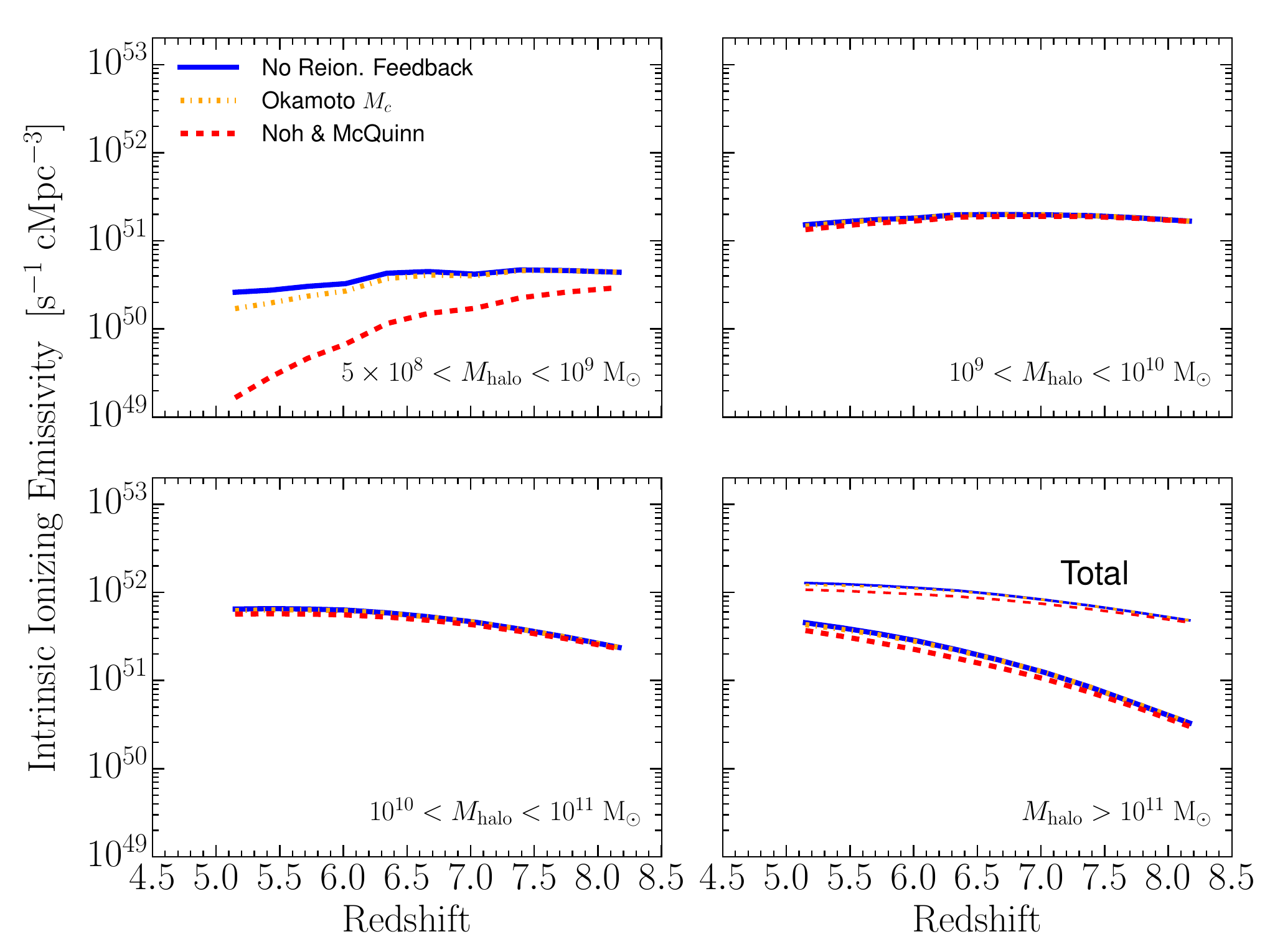}
    \caption{Same as in Figure \ref{fig:sfrd}, except here we plot the intrinsic ionizing emissivity of the galaxy population, i.e the rate of ionizing photon production within galaxies per unit volume. Reionization feedback can suppresses the intrinsic ionizing emissivity significantly in low-mass halos.  The effect in the \textsc{Noh \& McQuinn} model is significantly stronger than in the widely used \textsc{Okamoto $M_{\rm c}$ model.} }
    \label{fig:nion_int}
\end{figure}

\begin{figure}
    \centering
    \includegraphics[width=\columnwidth]{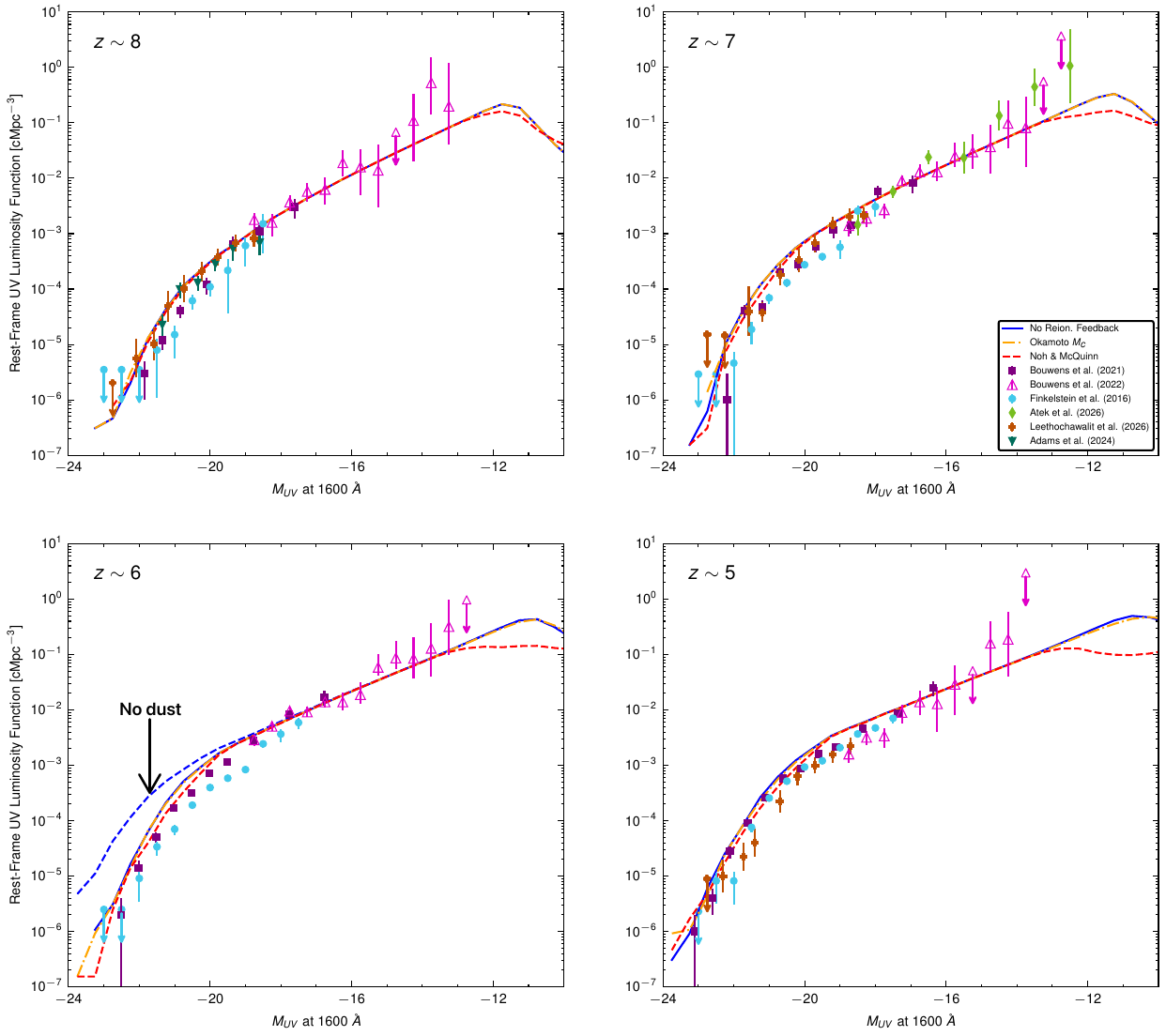}
    \caption{Rest-frame UV luminosity functions in our three reionization feedback models, compared to recent observational measurements.  The data points show a compilation of measurements as indicated in the plot legend \cite{2015ApJ...810...71F, 2021AJ....162...47B, 2022ApJ...940...55B, 2024ApJ...965..169A, 2026ApJ...998..217L, 2026OJAp....966816A}. These include measurements of the faint-end of the luminosity function using strong lensing magnification in the fields of foreground galaxy clusters. All model curves are attenuated by dust using the prescription of Ref \cite{2019MNRAS.483.2983Y}.  For reference, the blue/dashed curve in the lower-left panel shows the result in the \textsc{No Reion. Feedback} model before dust attentuation is applied. The faint-end rollover seen in all curves is a halo resolution effect from our minimum halo mass of $M_{\rm min} = 5\times10^8$ M$_\odot$.  In the \textsc{Noh \& McQuinn} model, reionization feedback suppresses the UV luminosity function at magnitudes fainter than $M_{\rm UV}\sim -13$, which are starting to be probed by observations of lensing clusters with JWST. }
    \label{fig:UVLF}
\end{figure}

\subsection{Effects of Reionization Feedback}
\label{sec:reion_feedback}

In what follows, we will distinguish between {\it intrinsic} and {\it escaped} ionizing emissivities, which we denote as $\dot{n}^{\rm int}_{\rm ion}$ and $\dot{n}^{\rm esc}_{\rm ion}$, respectively.  The former is the emissivity produced by the source population, while the latter subtracts off the fraction that is blocked by absorptions in the ISM and CGM. The latter is, of course, the quantity relevant for the Ly$\alpha$ forest transmission.  We begin by considering the effects of reionization feedback on the galaxy population.  Figure \ref{fig:sfrd} shows the star formation rate density (SFRD) in our semi-analytic models as a function of redshift near the end of reionization.  Each panel corresponds to a different halo mass bin, as noted in the bottom right corners. The blue/solid, orange/dot-dashed, and red/dashed curves show the \textsc{No Reion. Feedback}, \textsc{Okamoto $M_c$}, and \textsc{Noh \& McQuinn} models, respectively.   The thin set of curves at the top of the bottom-right panel show the total SFRD's for the entire galaxy population.  In the \textsc{Noh \& McQuinn} model, reionization strongly suppresses star formation in halos with mass below $M_{\rm halo} \sim 10^9$ M$_{\odot}$, as shown in the top-left panel.  The suppression effect builds with cosmic time as more halos are reionized and as the already-affected halos exhaust their remaining reservoirs of previously accreted gas. Reionization feedback has a small effect on the SFRD for halos more massive than $M_{\rm halo} \sim 10^9$ M$_\odot$. The $\lesssim 15$ \% difference in total SFRD between the \textsc{Noh \& McQuinn} and \textsc{No Reion. Feedback} models owes to the suppression effects in the former propagating up to massive halos through hierarchal assembly. {Although the IGM gas accretion rates for halos with masses $M_{\rm halo} \gtrsim 10^9$ M$_\odot$ are not affected by reionization feedback, these halos inherit diminished gas reservoirs from smaller progenitors that were suppressed by reionization in the past. This has a modest effect on the subsequent star formation rates of massive halos.}     

Notably, reionization feedback is much stronger in the more detailed \textsc{Noh \& McQuinn} model compared to the \textsc{Okamoto $M_c$} treatment used in previous studies.  As discussed in \cite{noh14}, this difference owes to the fact that $M_c$ significantly underestimates the halo mass scale below which accretion is suppressed by reionization.  The quantity $M_c$ is defined to be the mass-scale at which halos retain half of the cosmic baryon fraction, on average.  However, a halo that recently had its accretion shut off by reionization still retains its previously accreted gas for some time. Hence, there is a lag between the suppression mass scale and $M_c$.  Indeed, \cite{noh14} found that the former is significantly larger than the latter.  Applying a simple model based only on $M_c$ to a SAM, therefore, allows halos below the suppression scale to be accreting fresh gas when they should not be.  This leads to an overall underestimation of the effect of reionization feedback.        

Figure \ref{fig:nion_int} shows a similar comparison for the intrinsic ionizing emissivity. {We emphasize that the relationship between SFRD (Fig. \ref{fig:sfrd}) and $\dot{n}^{\rm int}_{\rm ion}$ is not assumed; for every galaxy $\dot{n}^{\rm int}_{\rm ion}$ is derived from its star formation history and stellar population synthesis model.} The results in Figure \ref{fig:nion_int} suggest that the escape fraction must be significantly larger for low-mass halos if reionization feedback plays a significant role in generating the emissivity decline -- a point we will return to later.  In the \textsc{No Reion. Feedback} model,  the intrinsic emissivity exhibits a decline in the $5 \times 10^8 < M_{\rm halo} < 10^9$~M$_{\odot}$ mass bin and a flat evolution in the $10^9 < M_{\rm halo} < 10^{10}$~M$_{\odot}$ mass bin, both without the effects of reionization feedback.   In Appendix \ref{app:feedback}, we investigate the physical effects that shape these curves.  We have performed variations of the \textsc{No Reion. Feedback} run, including one that turns off the effects of stellar feedback.  The evolution of the intrinsic emissivity in a given mass bin is set by a combination of hierarchical assembly and stellar feedback.  A competition between the declining accretion rates of fresh IGM gas with cosmic time and the growing number density of halos can determine whether the emissivity grows or declines. For instance, in the largest mass bin, $M_{\rm halo}>10^{11}$ M$_{\odot}$, we find that the halo abundance rises steeply enough to offset the overall declining accretion rates, driving growth in the intrinsic emissivity of that population. Stellar feedback blunts that growth somewhat, but not enough to reverse the evolution upwards. By contrast, in the lowest mass bin ($5 \times 10^8 < M_{\rm halo} < 10^9$~M$_{\odot}$), the growth of the halo abundance is insufficient to counter the declining accretion rates. In this mass bin, stellar feedback mostly affects the normalization of the intrinsic emissivity; the downward slope is shaped primarily by cosmological evolution in the gas accretion rate density.  

The suppression of star formation in the low-mass galaxies leads to a suppression at the faint end of the rest-frame UV luminosity function which, in principle, can be probed observationally.  Figure \ref{fig:UVLF} shows our model UV luminosity functions across the redshift range of interest here. The models are compared to a compilation of measurements from censuses of Hubble Space Telescope (HST) and JWST data \cite{2015ApJ...810...71F, 2021AJ....162...47B, 2022ApJ...940...55B, 2024ApJ...965..169A, 2026ApJ...998..217L, 2026OJAp....966816A}. These include measurements of the faint end of the luminosity function from cluster strong lensing fields.  As in the previous two figures, the curves show our three reionization feedback models. For reference, the blue/dashed curve in the lower-left panel ($z\sim6$) shows the luminosity function in the \textsc{No Reion. Feedback} model before applying the dust attenuation described in \S \ref{sec:SPS}.   In our models, dust has a significant effect in only the brightest galaxies with $M_{\rm UV} \lesssim -19$. The turnovers seen in all models at the faint end of the luminosity function, at magnitudes fainter than $M_{\rm UV} \approx -11.5$, owes to our minimum halo mass cutoff of $M_{\rm min} = 5\times 10^8$ M$_\odot$.  In the \textsc{Noh \& McQuinn} model, the effect of reionization feedback can also be seen at the faint end.  The suppression begins at magnitudes fainter than $M_{\rm UV}\approx -13$, brighter than the faintest magnitude probed by \cite{2026OJAp....966816A}.  However, they report no evidence for a turnover down to $M_{\rm UV} \lesssim -12.3$ at $z\sim 7$.  Our luminosity functions undershoot the measurement of \cite{2026OJAp....966816A} at $M_{\rm UV} \lesssim -14$, indicating that the models -- especially our implementation of \textsc{Noh \& McQuinn} -- predict a shallower faint-end slope. Qualitatively similar disagreements between these measurements and independent models are seen in Figures 10 and 11 of \cite{2026OJAp....966816A}.   Future determinations of the faint-end UV luminosity function from strong lensing fields will more tightly constrain the efficacy of reionization feedback.  These efforts should target lower redshifts, where the aggregate effects of IGM photoheating are stronger. Lastly, we note that if the faint-end slope is significantly steeper than in our models, reionization feedback may play a larger role in producing the emissivity decline than our models suggest owing to the larger number of small galaxies contributing to the ionizing photon budget.      

\subsection{Role of the Escape Fraction}

The results of the last section show that reionization feedback could play a role in producing a decline in the emissivity, provided that the ionizing photon budget at $z \lesssim 6 $ is dominated by low-mass galaxies.  However, the Ly$\alpha$ forest transmission probes the {\it escaped} ionizing emissivity, and we have so far only discussed the intrinsic emissivity. In this section, we explore the role that $f_{\rm esc}$ plays in the redshift evolution of the emissivity.

\subsubsection{Toy $f_{\rm esc}$ Models}
\label{sec:toymodels}

\begin{figure}
    \centering
    \includegraphics[width=0.95\columnwidth]{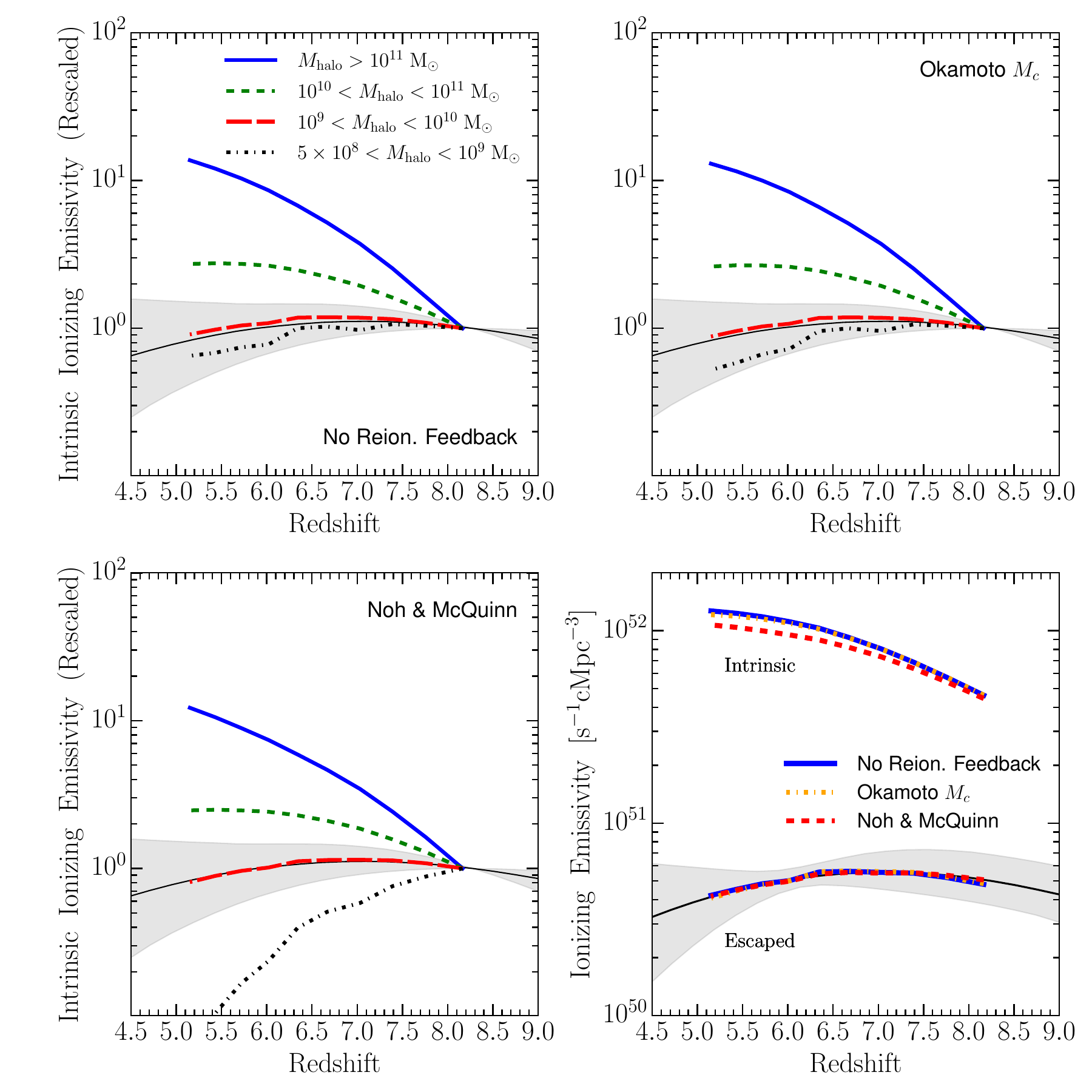}
    \caption{The redshift evolution of the intrinsic ionizing emissivity.  The top-left, top-right, and bottom-left panels correspond to the \textsc{No Reion. Feedback}, \textsc{Okamoto $M_{\rm c}$}, and \textsc{Noh \& McQuinn} reionization feedback models, respectively.  To compare the slope of the emissivity across different mass bins, each panel shows the intrinsic emissivity normalized to unity at $z\approx 8$. For qualitative comparison, the thin solid/black line in each panel shows the median emissivity model, also re-normalized, from the posterior of a joint CMB-forest analysis by \cite{2026arXiv261028765L} (see first paragraph of \S \ref{sec:toymodels}). The gray shading shows the model-dependent 95\% credible levels from \cite{2026arXiv261028765L}. The slope of the redshift evolution varies considerably with halo mass, with the lowest mass bins being strongly affected by stellar and reionization feedback.  {\bf Bottom-right panel:} The bottom set of curves (labeled ``escaped'') show the escaped ionizing emissivity assuming a toy model in which $f_{\rm esc}$ does not evolve with redshift. In this simplistic model, the escaped emissivity is a weighted sum of the different mass bins, where the weights are the emissivity-weighted mean escape fractions in the bins (eq. \ref{eq:nion_decomp}), obtained by a least squares fit to the median model from \cite{2026arXiv261028765L}.  {\it If the escape fraction does not evolve significantly with redshift}, recovering the emissivity decline requires that the escape fraction decrease strongly with halo mass, such that the escaped emissivity is dominated by low-mass halos.   }
    \label{fig:Nion_massbins_normalized}
\end{figure}

In this section, we develop some intuition with toy models of $f_{\rm esc}$. We begin with a simple model that fits the mean $f_{\rm esc}$ in 3--4 halo mass bins to the emissivity evolution measured by \cite{2026arXiv261028765L}.  Figure \ref{fig:Nion_massbins_normalized} motivates this exercise.  There, we compare the shapes of the emissivities in our models as a function of redshift when binned by halo mass.  To facilitate this comparison, all curves are normalized to unity at $z\approx 8.17$. The top-left, top-right, and bottom-left panels correspond to the \textsc{No Reion. Feedback}, \textsc{Okamoto $M_c$}, and \textsc{Noh \& McQuinn} models, respectively. In each panel, the different line styles correspond to the mass bins given in the plot legends.  For reference, the thin solid black curve and gray shading in each panel shows the median emissivity model and 95\% credible levels, also re-normalized, inferred by \cite{2026arXiv261028765L} in a joint CMB-forest analysis (see also Fig. \ref{fig1}). They constrain reionization models within the semi-analytic framework of \cite{2026arXiv260619449C} using IGM photoionization rate measurements from the Ly$\alpha$ forest \cite{Gaikwad2023} and {\it Planck} DR3 and DR4 TTTEEE likelihoods \cite{2021A&A...647A.128T, 2022PhRvD.105h3524T, 2020A&A...641A...5P}.  An important caveat is that their analysis adopts a fixed model for the IGM opacity to ionizing photons as well as a fixed cosmology. Here we use their constraints as a qualitative benchmark for the kinds of emissivity evolutions favored by forest transmission measurements.  The binned emissivities are broadly similar between the three reionization feedback models, except for the lowest mass bin, $5\times 10^8< M_{\rm halo} < 10^9$ M$_\odot$, in the \textsc{Noh \& McQuinn} model. Even in the \textsc{No Reion. Feedback} model, the emissivity in this bin, as well as in the $10^9< M_{\rm halo} < 10^{10}$ M$_\odot$ bin, declines with cosmic time due to falling accretion rates and a slower growth in the abundance of low-mass halos (see above discussion of Figure \ref{fig:nion_int}).  The decline of the lowest mass bin is significantly steeper when reionization feedback is included using the \textsc{Noh \& McQuinn} model.       

In all models, the redshift evolution of the $10^9< M_{\rm halo} < 10^{10}$ bin is notably similar to the median model from \cite{2026arXiv261028765L}. Because Figure \ref{fig:Nion_massbins_normalized} indicates that there are halo mass scales with different redshift evolutions for $\dot{n}^{\rm int}_{\rm ion}$, it is instructive to decompose the escaped emissivity into a linear combination of mass bins,

\begin{equation}
\dot{n}^{\rm esc}_{\rm ion}(z) = \sum_{i \in \mathrm{bins}} f_{\rm esc,i} \dot{n}^{\rm int}_{\rm ion, i}(z),
\label{eq:nion_decomp}
\end{equation}
where the $\dot{n}^{\rm int}_{\rm ion, i}(z)$ are the \textit{un-normalized} binned emissivities that appear in Figure \ref{fig:nion_int}, and $f_{\rm esc,i}$ is the $\dot{n}^{\rm int}_{\rm ion}$-weighted mean escape fraction in mass bin $i$, assumed here to be constant in redshift. (We will explore explicitly redshift-dependent $f_{\rm esc}$ models below.) Using the 10 SAM redshift bins, we perform a bounded ($ 0 \leq f_{\rm esc,i} \leq 1$) least-squares fit of equation (\ref{eq:nion_decomp}) to the median model from \cite{2026arXiv261028765L}, with equal weights for each redshift. For both the \textsc{No Reion. Feedback} and \textsc{Okamoto $M_c$} models, we combine the lowest two mass bins into a single bin because of their similar redshift evolutions.  For the \textsc{Noh \& McQuinn} model we use all 4 mass bins.  We note that, although the $f_{\rm esc,i}$ appearing in equation (\ref{eq:nion_decomp}) do not depend on redshift, the population averaged $f_{\rm esc}$ generally will because of the redshift dependencies of $\dot{n}^{\rm int}_{\rm ion, i}(z)$.

The best-fit solutions are the curves labeled ``Escaped'' in the bottom-right panel of Figure \ref{fig:Nion_massbins_normalized}.  The dashed curves in that panel show the corresponding intrinsic emissivities.  From lowest to highest mass bin, the best-fit $f_{\rm esc,i}$ values are $(0.25, 0,0 )$, $(0.25, 0, 0)$, and $(0.75, 0.18, 0.03, 0)$ for the \textsc{No Reion. Feedback}, \textsc{Okamoto $M_c$}, and \textsc{Noh \& McQuinn} models, respectively. Notably, the models require $f_{\rm esc,i}$ to be suppressed for $M_{\rm halo}\gtrsim 10^{10}$ M$_\odot$, irrespective of the degree of reionization feedback.  Although a decrease in $f_{\rm esc}$ with halo mass is reported in some studies based on cosmological simulations \cite[e.g.][]{2022MNRAS.515.2386R,2020MNRAS.498.2001M}, the predicted mass dependence varies considerably between simulations, with other studies reporting more complex or non-monotonic relationships  \cite[e.g.][]{2023MNRAS.520.2757Y,2023MNRAS.521.3077K}. It is unclear what physical mechanism would result in the sharp suppression at $M_{\rm halo}\gtrsim 10^{10}$ M$_\odot$ implied by this illustrative analysis.  While we caution against over-interpreting the particular best-fit $f^{\rm eff}_{\rm esc,i}$ values, this simplistic exercise demonstrates two key points: (1) Assuming that the escape fraction at fixed halo mass does not evolve with redshift, the escaped ionizing emissivity must be dominated by smaller halos with masses $M\lesssim 10^{10}$~M$_{\odot}$ in order to produce a decline; (2) Reionization feedback may not be required to produce the observed emissivity decline. {A combination of declining accretion rates, slower growth in the abundance of low-mass halos, and stellar feedback may be sufficient to achieve the effect (see \S \ref{sec:reion_feedback} and Appendix \ref{app:feedback}).} This is demonstrated by the fact that the median model of \cite{2026arXiv261028765L} can be recovered equally well in the \textsc{No Reion. Feedback} model.

We emphasize that the solutions in Figure \ref{fig:Nion_massbins_normalized} are not unique.  Indeed, we can achieve an equally good fit with another simplistic model in which $f_{\rm esc}$ depends only on redshift explicitly (and not on halo mass) as $f_{\rm esc} = f_6 \left[ (1+z)/(1+6)\right]^{\beta}$, where $f_6$ is a constant representing the escape fraction at $z=6$. In this case, the best fit values are $(f_6,\beta)= (0.046, 3.1)$, $(0.047,3.0)$, and $(0.053, 2.7)$ for the \textsc{No Reion. Feedback}, \textsc{Okamoto $M_c$}, and \textsc{Noh \& McQuinn} models, respectively. These yield similar residual RMS values to the $z$-independent models shown in the bottom-right panel of Figure \ref{fig:Nion_massbins_normalized}. (Because they fall on top of each other, we omit the power-law fits from Fig. \ref{fig:Nion_massbins_normalized} for clarity.)  This exercise shows that, in models where $f_{\rm esc}$ is independent of the halo mass, the escape fraction must decline steeply with time in order to generate the emissivity decline. 

The results from these toy models highlight the fact that the Ly$\alpha$ forest transmission alone cannot be used to disentangle the role of the escape fraction in generating the emissivity decline.  Lacking a clean empirical route for attacking this problem, we can make further progress by appealing to state-of-the-art simulations. 

\subsubsection{Physical Model Based on Simulation Results}

The escape fraction is notoriously difficult to model in cosmological simulations because it is shaped by uncertain interstellar medium (ISM) physics on short distance and time scales ($\sim 1$ pc and few Myr, respectively), thus requiring an enormous dynamic range. The most recent simulations, however, agree broadly on the physical conditions associated with ionizing photon leakage from galaxies \cite[e.g.][]{2020MNRAS.498.2001M, 2022MNRAS.515.2386R, 2023MNRAS.518..270K, 2023MNRAS.520.2757Y, 2023MNRAS.521.3077K, 2024MNRAS.529.3751C,2026arXiv260614671S, 2026arXiv261115931C}.  Leakage is thought to occur through channels cleared of neutral hydrogen by stellar feedback in the form of supernovae, winds, and photoionization, making the phenomenon highly variable/stochastic in time on a galaxy-by-galaxy basis. Independent simulations also agree that leakage peaks in galaxies of low/intermediate masses undergoing vigorous star formation (with commensurate feedback activity), since these possess shallower neutral gas channels to clear compared to their more massive/evolved counterparts.{  In more massive/enriched galaxies, dust attenuation may play a role in lowering $f_{\rm esc}$ \cite{2020MNRAS.498.2001M, 2023MNRAS.520.2757Y} (see however \cite{2020arXiv200105508Y, 2022MNRAS.515.2386R}).  Radiation pressure on dust could also facilitate the creation low-column-density channels in dense star-forming clouds \cite{Ferrara2025, 2025ApJ...987...12M}.} 

Interestingly, in their analysis of the \textsc{SPHINX} simulation \cite{2018MNRAS.479..994R}, Ref. \cite{2022MNRAS.515.2386R} found that the $f_{\rm esc}$ of \textsc{SPHINX} galaxies decrease steeply, on average, with cosmic time. The evolution is qualitatively consistent with the trend earlier noted in the \textsc{FIRE-2} simulations \cite{2020MNRAS.498.2001M}.  Ref. \cite{2022MNRAS.515.2386R} proposed an intuitive root cause for the decrease.   They found that $f_{\rm esc}$ correlates strongly with the {\it maximum value} of the specific star formation rate (sSFR) over the past $\sim 50$ Myr, essentially a proxy for the presence and intensity of recent starbursts, and associated feedback, relative to the size of the system.  Specifically, they defined

\begin{equation}
\mathrm{sSFR}_{10} = \frac{M_*(\mathrm{age}<10\mathrm{Myr})~/~10\mathrm{Myr}}{{M_*}}
\label{eq:sSFR}
\end{equation}
The numerator contains the mass in stars with ages less than 10 Myr, while the denominator is the total stellar mass of the galaxy. Then, $\mathrm{sSFR}_{10,{\mathrm{max}}}$ is the maximum among $\mathrm{sSFR}_{10}$ values computed in 10 Myr windows out to a lookback time of 50 Myr.\footnote{Ref. \cite{2022MNRAS.515.2386R} noted that their results are insensitive to the exact choice of lookback time, provided that it is much larger than 10 Myr.}  Ref. \cite{2022MNRAS.515.2386R} found a correlation between $f_{\rm esc}$ and $\mathrm{sSFR}_{10,{\mathrm{max}}}$ that evolves only weakly with redshift.  However, on average, $\mathrm{sSFR}_{10,{\mathrm{max}}}$ decreases steeply with cosmic time because: (1) Accretion and merger rates of the galaxy population evolve quickly as the Universe expands, resulting in less burstiness in the star formation; (2) Stellar masses (denominator of eq. \ref{eq:sSFR}) grow quickly.  Ref. \cite{2022MNRAS.515.2386R} proposed that these effects yield a rapid decline in the population-averaged $f_{\rm esc}$. We will refer to this picture as the \textsc{SPHINX} model for $f_{\rm esc}$. Followup studies expanded on these findings by considering the dependence of $f_{\rm esc}$ on various physical quantities \cite{2022A&A...663A..66M, 2023MNRAS.518..270K}, and formulating the model in terms of directly observable spectral features \cite{2024MNRAS.529.3751C}.

Conveniently, Table D1 of \cite{2022MNRAS.515.2386R} provides fitting functions describing the correlation between LyC luminosity-weighted mean $f_{\rm esc}$ and $\mathrm{sSFR}_{10,{\mathrm{max}}}$, measured from the SPHINX simulation. We implement the \textsc{SPHINX} model in our calculations using the star formation histories generated by our SAM runs. Because we assign the LyC luminosity-weighted mean $f_{\rm esc}$ from \textsc{SPHINX} to each galaxy at fixed $\mathrm{sSFR}_{10,{\mathrm{max}}}$, this approach does not fully capture the substantial scatter in $f_{\rm esc}$ within the galaxy population. Ref. \cite{2025arXiv251107543W} developed an alternative stochastic prescription by fitting the distribution of $f_{\rm esc}$ for \textsc{SPHINX} galaxies at fixed sSFR$_{10,\rm max}$, and drawing randomly from this distribution.  While their prescription is useful for modeling the scatter in $f_{\rm esc}$, the random drawing does not preserve the correlation between $f_{\rm esc}$ and $\dot{n}^{\rm int}_{\rm ion}$ at fixed sSFR$_{10,\rm max}$. We believe the prescription of \cite{2022MNRAS.515.2386R} is therefore better-suited for modeling the global emissivity evolution. Below, we will find that the \textsc{SPHINX} prescription captures the redshift evolution of the escaped emissivity well, but under-predicts the normalization by a factor of $\approx 4.5$.  We therefore multiply all $f_{\rm esc}$ values by a factor of 4.5, as we are mainly concerned with modeling the steep redshift evolution. This procedure results in $\dot{n}^{\rm int}_{\rm ion}$-weighted average $f_{\rm esc}$ ranging from 3--$12$\% between $z=5-8$. We note that Ref. \cite{2025arXiv251107543W} also found it necessary to rescale their $f_{\rm esc}$ values by a factor of 2 to avoid reionization ending too late in their models. {These values are broadly consistent with recently measured values from JWST observations, within large uncertainties. Based on joint fitting of photometry and NIRSpec spectra, Ref \cite{Papovich2025} found mean escape fractions of $6.5 \pm 2.2\%$ at $4.5 < z < 6$ and $2.6 \pm 1.4\%$
at $6 < z < 9$. Ref. \cite{Giovinazzo2026} used ``picket-fence'' modeling of NIRSpec spectra at $5 < z < 10$ to infer $f_{\rm esc}\sim 10~\%$.  Another approach based on empirical indicators of ionizing photon leakage yields $f_{\rm esc}\simeq 13~\%$ $6 < z < 9$ \cite{Mascia2024}.} 

\begin{figure}
    \centering
    \includegraphics[width=\columnwidth]{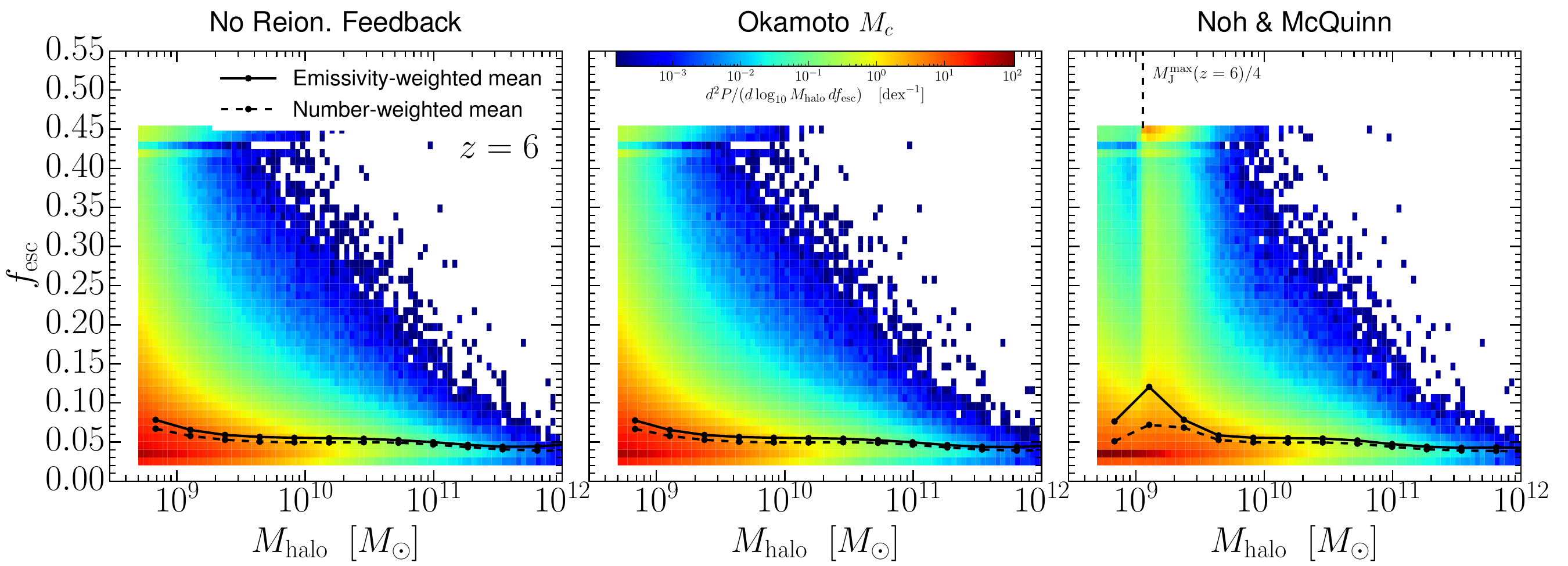}
    \includegraphics[width=\columnwidth]{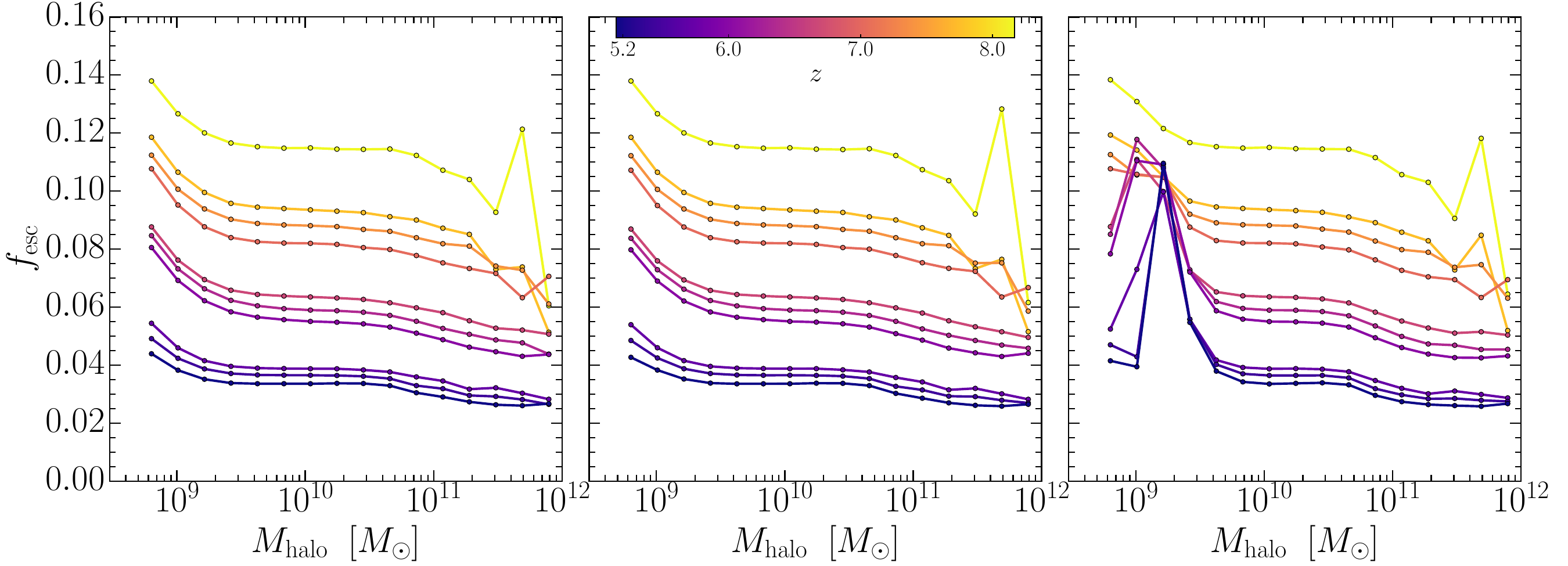}
    \caption{Escape fractions in our SAM implementation of a model based on results from the \textsc{SPHINX} simulation, from Ref. \cite{2022MNRAS.515.2386R}. The model is based on a correlation measured in \textsc{SPHINX} between $f_{\rm esc}$ and the maximum specific star formation rate within a lookback time of tens of Myr, essentially a proxy for the presence and intensity of recent starbursts, and associated feedback, relative to the size of the system. All $f_{\rm esc}$ are scaled up by a constant factor of 4 relative to the original model in \cite{2022MNRAS.515.2386R} for better agreement with the ionizing emissivities inferred from Ly$\alpha$ forest transmission measurements. The three columns correspond to the reionization feedback models, as annotated at the top. {\bf Top row:} The distribution of $f_{\rm esc}$ values as a function of halo mass. The solid and dashed curves show the $\dot{n}^{\rm int}_{\rm ion}$--weighted mean $f_{\rm esc}$ and number-weighted average, respectively, with the former being the relevant quantity for this study (see footnote \ref{footnote:fesc}). Horizontal bands and cutoffs in these distributions are artifacts of our numerical implementation of the model (see main text).  The vertical features seen in the \textsc{Noh \& McQuinn} model result from an interplay between reionization feedback and $f_{\rm esc}$. {\bf Bottom row:} Redshift evolution of the escape fraction. We show the $\dot{n}^{\rm int}_{\rm ion}$--weighted mean $f_{\rm esc}$ at 10 redshifts between $z\approx 8 - 5$, with redshift decreasing in the downward direction.  In the \textsc{SPHINX} model, the decreasing specific star formation rates of the galaxy population yield a steep downward evolution in $f_{\rm esc}$ values \cite{2022MNRAS.515.2386R}.}
    \label{fig:fesc_heatmap}
\end{figure}

The top row of Figure \ref{fig:fesc_heatmap} shows the joint distribution of $f_{\rm esc}$ and $M_{\rm halo}$ values for the three reionization feedback models at $z=6$. The solid overlaid curves show the $\dot{n}^{\mathrm{int}}_{\rm ion}$-weighted mean $f_{\rm esc}$ in bins of $M_{\rm halo}$, while the dashed curves show the number-weighted mean.  We note that the $\dot{n}^{\mathrm{int}}_{\rm ion}$-weighted mean is the more relevant quantity for our purposes; in a given $M_{\rm halo}$ bin, it gives the $f_{\rm esc}$ value that converts the intrinsic emissivity to the escaped emissivity.\footnote{\label{footnote:fesc}This can be seen by writing the total escaped emissivity as $\dot{n}^{\rm esc}_{\rm ion} = \sum_i f_{\mathrm{esc},i}~\dot{n}^{\rm int}_{\mathrm{ion},i}$, where $f_{\mathrm{esc},i}$ and $\dot{n}^{\rm int}_{\mathrm{ion},i}$ are, respectively, the escape fraction and intrinsic ionizing emissivity contribution of galaxy $i$ (the ionizing photon production rate of the galaxy divided by the simulation volume). If we bin the galaxies by halo mass, we can write this sum as $\dot{n}^{\rm esc}_{\rm ion} = \sum_j \sum_{i\in j} f_{\mathrm{esc},i}~\dot{n}^{\rm int}_{\mathrm{ion},i}$, where bin $j$ has central halo mass value $M_j$. Defining the intrinsic ionizing emissivity-weighted mean escape fraction to be $\langle f_{\rm esc}\rangle_{\dot{n}}(M_j) \equiv \sum_{i\in j} f_{\mathrm{esc},i}~\dot{n}^{\rm int}_{\mathrm{ion},i}~/~\sum_{i\in j}\dot{n}^{\rm int}_{\rm ion,i}$, we can write $\dot{n}^{\rm esc}_{\rm ion}= \sum_j \langle f_{\rm esc}\rangle_{\dot{n}}(M_j)~\left(\sum_{i\in j} \dot{n}^{\rm int}_{\rm ion,i}\right)$. So $\langle f_{\rm esc}\rangle_{\dot{n}}(M_j)$ is the escape fraction that converts the intrinsic emissivity in bin $j$ into the corresponding escaped emissivity.}   Our implementation of the \textsc{SPHINX} $f_{\rm esc}$ prescription results in a few un-physical features seen as the horizontal stripes in the heat maps. These arise because we clip $x\equiv \log_{10}\left( \mathrm{sSFR}_{10,\mathrm{max}}/\mathrm{Gyr}^{-1}\right)$ to be within the ranges of the fits in Table D1 of \cite{2022MNRAS.515.2386R}, which imposes upper and lower limits in $f_{\rm esc}$ and produces the other sharp horizontal features.  Consider first the \textsc{No Reion. Feedback} model (left panel).  The escape fraction exhibits a large scatter at fixed $M_{\rm halo}$ and the mean value decreases with halo mass.  The $\dot{n}^{\rm int}_{\rm ion}$-weighted mean decreases by a factor of $\approx 2$ over the mass range.  The \textsc{Okamoto $M_c$} model exhibits similar behavior. The effect of reionization feedback is seen more starkly in the \textsc{Noh \& McQuinn} model. There we see a sharp vertical feature corresponding to the mass scale below which accretion is shut off in the model.  For reference, we show $M^{\rm max}_{\rm J}/4$ as the vertical dashed line segment near the top of the plot. Although the vast majority of collapsed halos at $z=6$ are reionized at $z_{\rm re} < z_{\rm ta}=10.1$, this scale roughly marks the suppression scale within the model (see Fig. \ref{fig:Mchar}).   The node of large $f_{\rm esc}$ occurring at $M_{\rm halo}$ values just above this boundary corresponds to previously gas-starved halos with large sSFR's owing to the recently reactivated accretion.

\begin{figure}
    \centering
    \includegraphics[scale=0.5]{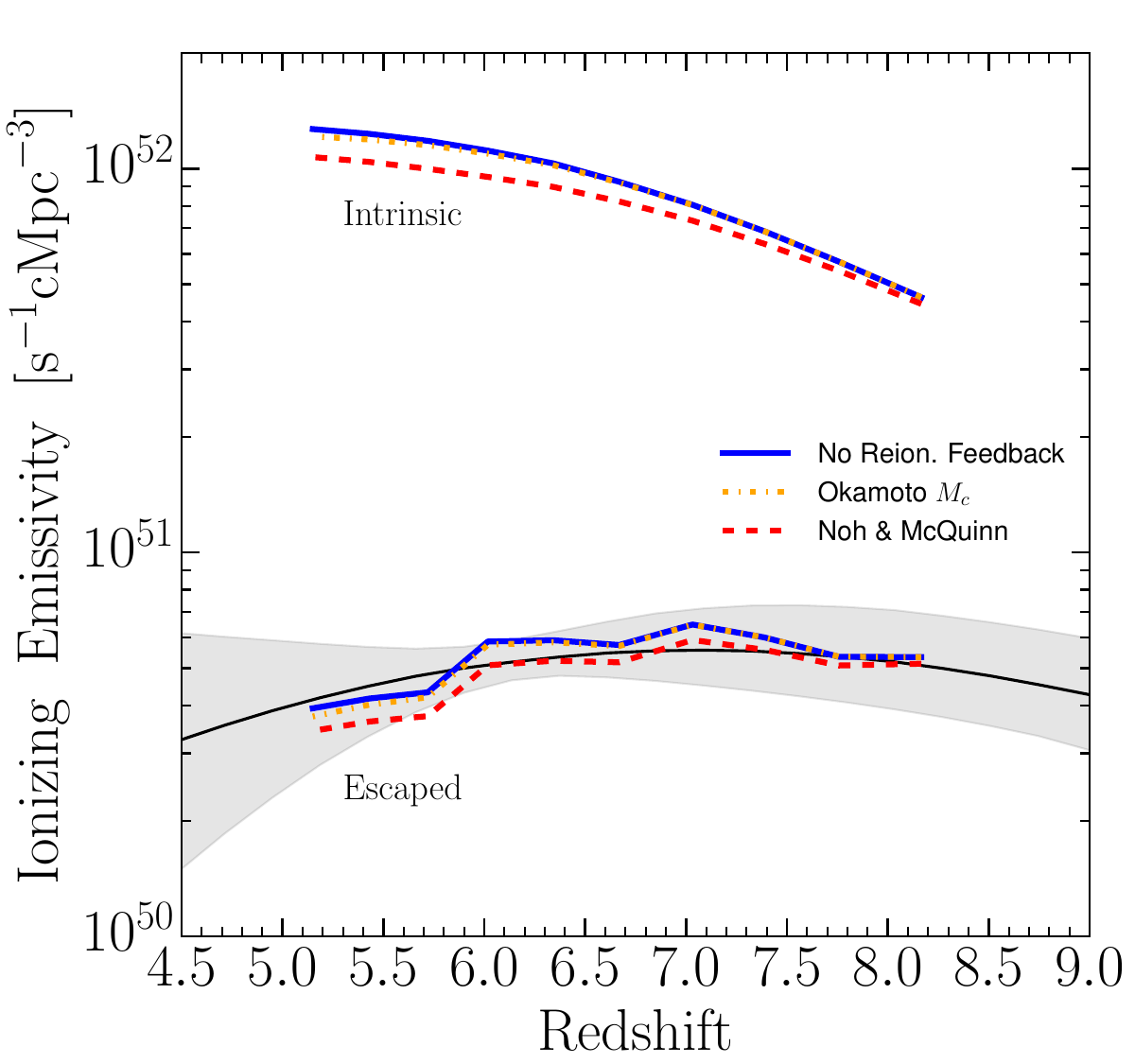}
    \caption{Evolution of the escaped ionizing emissivity in our implementation of the \textsc{SPHINX} model for $f_{\rm esc}$. For reference, the top set of curves show the intrinsic emissivities (i.e. without $f_{\rm esc}$ applied).  The black/solid curve and gray shading correspond to the median and 95\% credible levels from the joint CMB-forest analysis of \cite{2026arXiv261028765L}. These constraints assume a fixed model for the IGM opacity to ionizing photons, as well as a fixed cosmology. We use them here for qualitative comparison to the redshift evolutions in our SAM models. In these models, reionization feedback has little effect on the redshift evolution of the emissivity because it mainly affects halos with mass $M_{\rm halo}\lesssim 10^9$ M$_{\odot}$. In the \textsc{SPHINX} model, a strong decline in $f_{\rm esc}$ arises from the diminishing efficacy of stellar feedback to create low-column channels through the ISM.  This effect naturally yields an emissivity redshift evolution that agrees qualitatively with Ly$\alpha$ forest measurements. We emphasize that the only ``tuning'' here is a global rescaling of $f_{\rm esc}$ by a factor of 4.5 to set the normalizations of these curves, resulting in $\dot{n}^{\rm int}_{\rm ion}$-weighted average $f_{\rm esc}$ ranging from 3--$12$\% between $z=5-8$.}
    \label{fig:nion_sphinx}
\end{figure}

The bottom row of Figure \ref{fig:fesc_heatmap} shows the redshift evolution of the $\dot{n}^{\rm int}_{\rm ion}$-weighted mean $f_{\rm esc}$.  In each panel, from top to bottom, the curves correspond to $z=8.18$, $7.76$, $7.39$, $7.03$, $6.67$, $6.34$, $6.02$, $5.72$, $5.43$, and $5.15$.  The clustering together of curves into groups of 3 owes again to an implementation detail of the \textsc{SPHINX} model.  Ref. \cite{2022MNRAS.515.2386R} provided fitting function parameters for several redshift bins. Three curves clustered together, for example, utilize the same set of fit parameters.\footnote{{Specifically, the fit for the LyC luminosity-weighted average $f_{\rm esc}$ in a given sSFR$_{10}$ bin is given in terms of a 3rd order polynomial, $f_{\rm esc} = 10^{a+bx+c x^2 +d x^3}$, where $x=\log(\mathrm{sSFR}_{10}/\mathrm{Gyr}^{-1})$. The parameters $a$, $b$, $c$, and $d$ are provided in redshift bins, e.g. $z=$9--8, 8--7, 7--6, 6--4.7.}}  The trend of $f_{\rm esc}$ decreasing with $M_{\rm halo}$, noted above, appears somewhat stronger at higher redshift.  The reionization feedback feature in the \textsc{Noh \& McQuinn} model strengthens with time as more halos are reionized, and as more halos grow to masses above the suppression threshold. The most striking feature of these panels, however, is the strong redshift evolution.  Between $z=8.18$ and $z=5.15$, the average $f_{\rm esc}$ decreases by a factor of $\approx 3-2.5$ across all models. 

Figure \ref{fig:nion_sphinx} is the central result of this paper. There we show the total escaped ionizing emissivity using the \textsc{SPHINX} $f_{\rm esc}$ prescription with our 3 reionization feedback models.  For reference, the top set of curves shows the corresponding intrinsic emissivities (without $f_{\rm esc}$ applied).  The $\dot{n}^{\rm esc}_{\rm ion}(z)$ in our models is consistent with the downward evolution required by Ly$\alpha$ forest measurements, compared for example to the results of \cite{2026arXiv261028765L}, again shown as the thin curve and gray shading.  Note that the overall normalization of the model curves have been adjusted by boosting all $f_{\rm esc}$ in the \textsc{SPHINX} prescription by a factor of 4.5.  Our results nevertheless show that a declining emissivity is expected based on the results of radiative hydrodynamics simulations. Reionization feedback is believed to be an essential regulator of galaxy formation in the lowest-mass halos \cite{1992MNRAS.256P..43E,2000ApJ...539..517B,2008MNRAS.390..920O}, and is an important ingredient for understanding the abundance and stellar populations of Local Group ultra-faint dwarfs \cite{2000ApJ...539..517B,2002ApJ...572L..23S, 2014ApJ...796...91B,2019ARA&A..57..375S}.  However, the similarity in the shapes of $\dot{n}^{\rm esc}_{\rm ion}(z)$ in Figure \ref{fig:nion_sphinx} shows that reionization feedback does not play a significant role in generating the emissivity decline within our models.    

We have argued that a declining emissivity at $z\sim6$ requires an escape fraction that decreases with increasing halo mass, decreases explicitly with cosmic time, or some combination of the two.  The \textsc{SPHINX} $f_{\rm esc}$ model of Ref. \cite{2022MNRAS.515.2386R}, calibrated with a state-of-the-art radiative hydrodynamics simulation, exhibits both of these features. Within that model, however, the declining emissivity is a result of the decreasing intensity of recent starbursts and an associated weakening in the efficacy of its feedback in producing LyC leakage. {Analytic models provide physical insight into this weakening. In the dense star-forming regions of high-$z$ galaxies, gas collapses on time scales shorter than those over which supernova feedback disrupts them, resulting in the buildup of more violent starbursts.  At lower redshift, the collapse time scales become longer than the disruption time scales, allowing supernova to more effectively regulate the starburst intensity. At the same time, as halos grow, the gas becomes less prone to expulsion; star formation settles down into a steadier state \cite{2018MNRAS.473.3717F, 2022MNRAS.511.3895F}.}  In this picture, the timing of the decline during the final stages of reionization is largely coincidental and not tied to reionization feedback.

\section{Discussion}
\label{sec:discussion}

\begin{figure}
    \centering
    \resizebox{7.6cm}{!}{\includegraphics[]{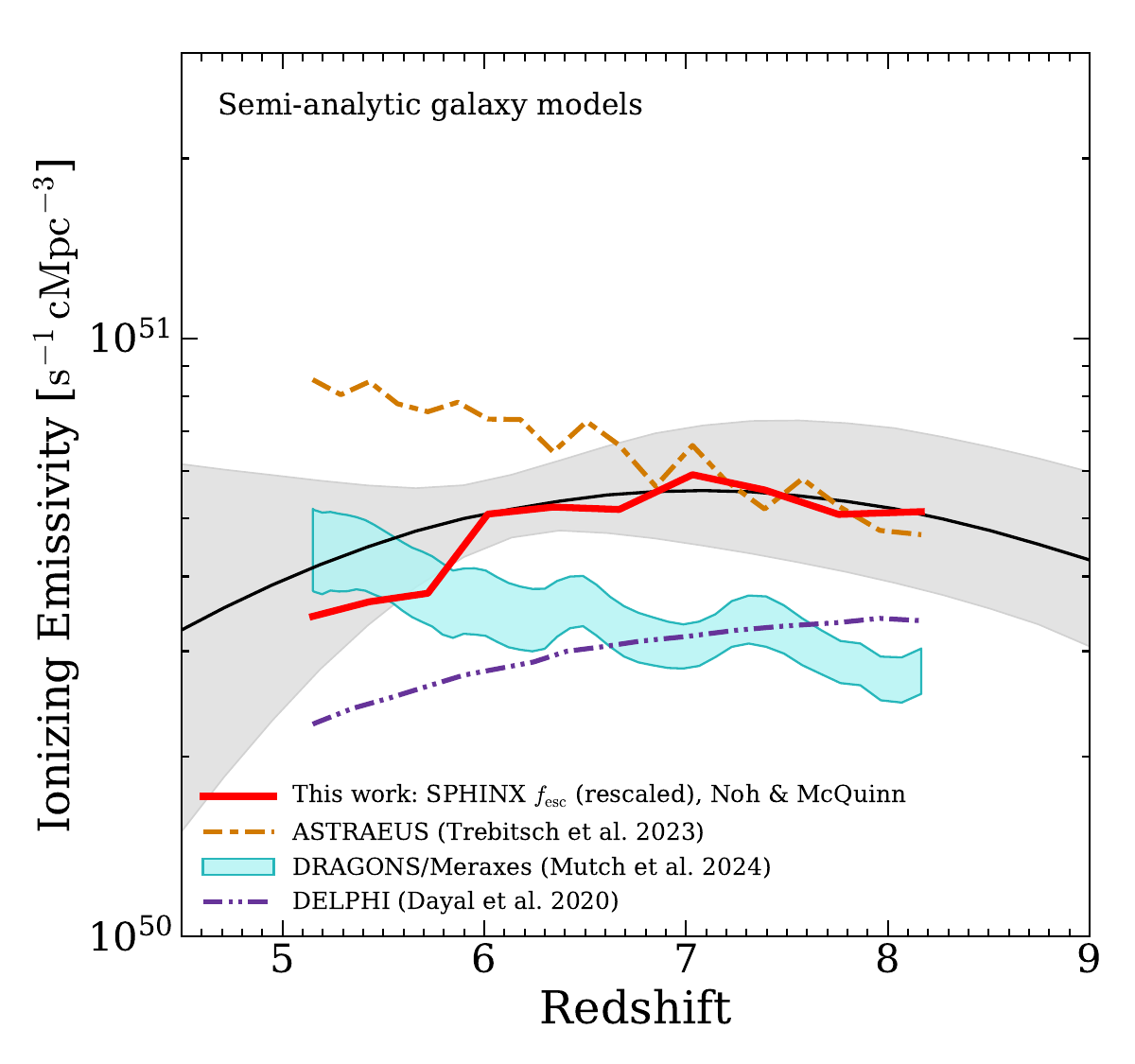}}
 \resizebox{7.6cm}{!}{\includegraphics[]{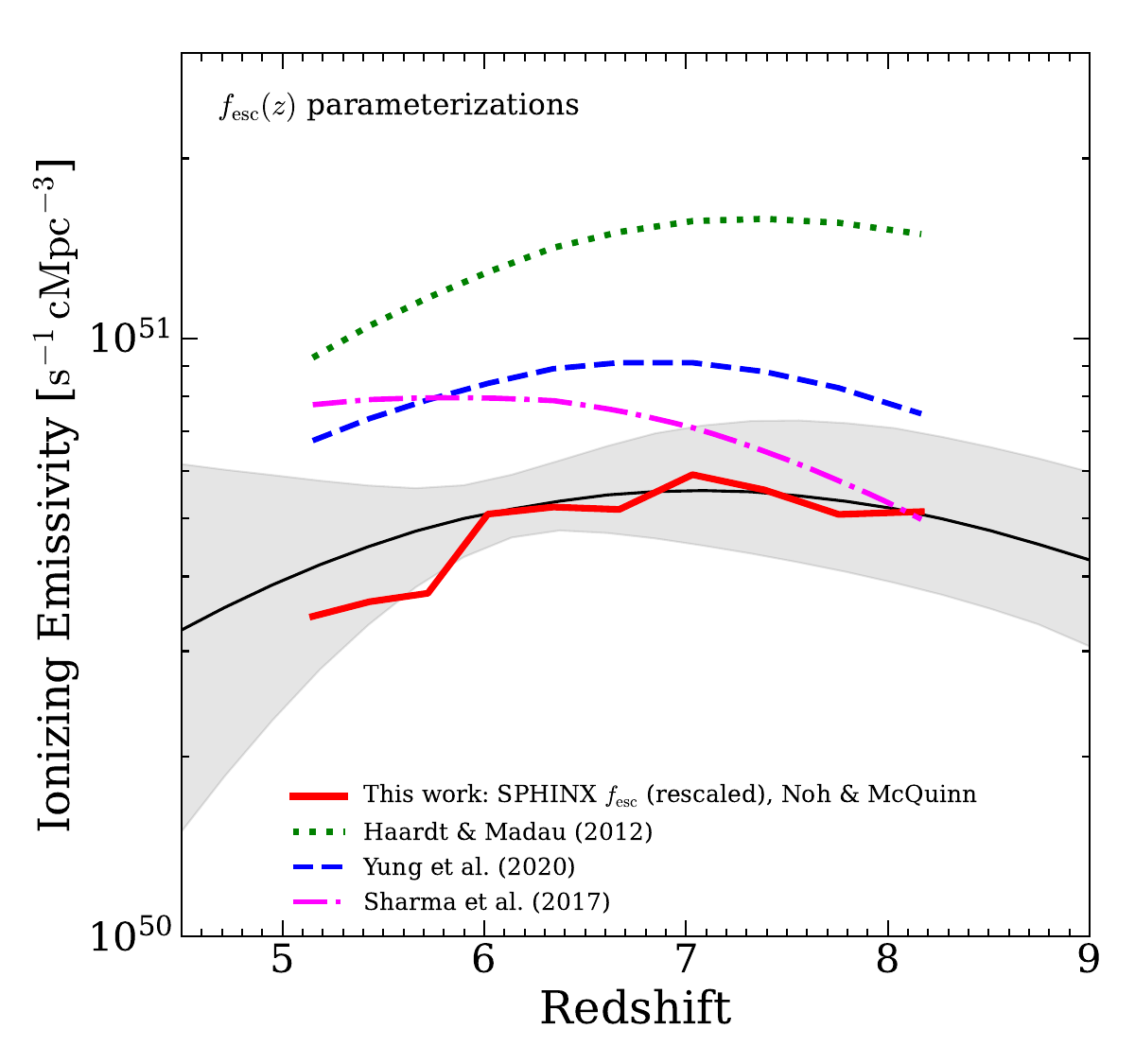}}
    \caption{Comparison of the escaped ionizing emissivity in our model to previous works. In both panels, the red/solid curve shows our full model including the rescaled \textsc{SPHINX} $f_{\rm esc}$ and \textsc{Noh \& McQuinn} reionization feedback prescriptions.  As in other figures, the black/solid curve and gray shading correspond to the median and 95\% credible levels from the joint CMB-forest analysis of \cite{2026arXiv261028765L}, used here as a qualitative benchmark. {\bf Left:} Ionizing emissivities in 3 other SAM models. {\bf Right:} A comparison to simple parameterizations for $f_{\rm esc}(z)$, which are summarized in Table \ref{tab:discussion_models}.  The curves show the results of multiplying these parameterizations with the intrinsic ionizing emissivity from our model. In all cases, the important comparison is with the redshift-evolution of the emissivity -- not the normalization (as we have rescaled the normalization of the \textsc{SPHINX} model for qualitative agreement with the solid black curve).   }
    \label{fig:comparisons}
\end{figure}

In this section, we discuss our findings in the context of previously published results.  In the left panel of Figure \ref{fig:comparisons}, we compare our full model (including the rescaled \textsc{SPHINX} $f_{\rm esc}$ and the \textsc{Noh \& McQuinn} reionization feedback prescriptions) to independent results from 3 other SAM frameworks.  The \textsc{ASTRAEUS} model is based on halo merger trees from \textsc{VSMDPL} (the same cosmological simulation used in the current paper), coupled to a spatially varying ionizing background model from a semi-numeric reionization code \cite{trebitsch23}. In their model, $f_{\rm esc}$ is proportional to the fraction of gas that is expelled by supernovae at each time step, with a maximum value of $f_{\rm esc}=0.24$.  Although their model includes a contribution from AGN to reionization, Figure  \ref{fig:comparisons} shows only the stellar contribution for comparison to our results.  \textsc{DRAGONS/Meraxes} also couples a SAM to a semi-numeric reionization simulation \cite{mutch24}.  The escape fraction is modeled with a four-parameter function of halo mass, with no explicit redshift dependence.  These parameters are constrained in a joint fit of \textit{Planck} 2016 $\tau_{\rm es}$ \cite{2016arXiv160503507P}, Ly$\alpha$ forest $\dot{n}^{\rm esc}_{\rm ion}$ measurements \cite{daloisio-gal}, and IGM neutral fraction constraints \cite{mcgreer15, 2017MNRAS.466.4239G,2018arXiv180206066D,2019MNRAS.484.5094G,2020ApJ...896...23W, 2018ApJ...856....2M,2019ApJ...878...12H}. The shaded region shows the 68 \% credible interval from their fit.  Lastly, we show the stellar-only \textsc{DELPHI} model, which adopts an explicitly redshift-dependent escape fraction, $f_{\rm esc} = 0.02 [ (1+z)/7 ]^{2.8}$.  This model was calibrated jointly to \textit{Planck} 2018 $\tau_{\rm es}$ \cite{2020A&A...641A...6P} and Ly$\alpha$ forest estimates of $\dot{n}^{\rm esc}_{\rm ion}$ at $z\simeq4$--6 from the model of \cite{kuhlen12}. Of the three models compared in Figure \ref{fig:comparisons}, the stellar-only \textsc{DELPHI} model exhibits a redshift evolution similar to our model. This agreement owes to the calibration of their assumed functional form of $f_{\rm esc}(z)$ to forest measurements.  

The right panel of Figure \ref{fig:comparisons} compares our model against a number of parameterizations for $f_{\rm esc}(z)$ in the literature.  In each case, the functional form, as well as a brief description of each model, are provided in Table \ref{tab:discussion_models}.  The corresponding curves in Figure \ref{fig:comparisons} were obtained by multiplying the parameterizations by the $\dot{n}^{\rm int}_{\rm ion}(z)$ in our full model (red/dashed curve in Fig. \ref{fig:nion_sphinx}).  A steeply declining $f_{\rm esc}$, similar to our findings, has long been assumed in synthesis models of the ionizing radiation background, such as the widely used model of \cite{haardt12} (green/dotted). In these models, the $f_{\rm esc}$ ansatz was motivated by the combination of Ly$\alpha$ forest $\dot{n}^{\rm esc}_{\rm ion}$ measurements and the need to complete reionization at an early-enough epoch to satisfy Gunn-Peterson and CMB $\tau_{\rm es}$ constraints. Additional examples (not shown) include the ionizing background models of \cite{fauchergiguere20} and \cite{puchwein19}, for which $f_{\rm esc}$ is assumed to scale as $(1+z)^{2.5}$ and $(1+z)^{3.97}$, respectively. 

The right panel of Figure \ref{fig:comparisons} also shows the model of \cite{2020MNRAS.tmp.1918Y} (blue/dashed), which was obtained using the Santa Cruz SAM. They performed a Bayesian MCMC analysis to jointly constrain a parametric form for $f_{\rm esc}(z)$ and the strength of stellar feedback using the $\dot{n}_{\rm ion}^{\rm esc}$ measurements of \cite{becker13} and \textit{Planck} 2016 $\tau_{\rm es}$ \cite{2016arXiv160503507P}. The parametric form in Table \ref{tab:discussion_models} adopts the median parameter values of their marginalized posteriors.  The model of \cite{sharma17} (magenta/dot-dashed) is based on star forming galaxies in the \textsc{EAGLE} simulation. They assign star formation rate surface densities ($\Sigma_{\rm SFR}$) to young star particles in the simulation and assume that 20\% of their ionizing photons escape the galaxy if $\Sigma_{\rm SFR} \geq 0.1$ M$_\odot$ yr$^{-1}$ kpc$^{-2}$. Zero photons escape otherwise. This $\Sigma_{\rm SFR}$ threshold was motivated by observations of local starbursts. Ref. \cite{sharma17} then fit the non-ionizing UV luminosity-weighted averaged $f_{\rm esc}$ in their model with a simple power law in $(1+z)$.  We apply this average to our $\dot{n}^{\rm int}_{\rm esc}(z)$ as an approximation.  Within the Santa Cruz SAM framework, Ref. \cite{2020MNRAS.494.1002Y} found weak evolution in $\xi_{\rm ion}$ at fixed UV luminosity, suggesting that this approximation should give a reasonable sense for the redshift-evolution predicted by the model of \cite{sharma17}.  While their model produces a decline in $f_{\rm esc}$ with cosmic time, it is not steep enough to counter the rising $n^{\rm int}_{\rm ion}(z)$ in our model (this is also supported by their Fig. 7). The redshift evolutions in the other two models are broadly consistent with our result.  Again, the agreement owes to the assumed or fitted redshift-dependence of $f_{\rm esc}$.  By contrast, the redshift dependence is not tuned in our implementation of the \textsc{SPHINX} model; it arises naturally from the star formation histories of the SAM galaxies.  Another interesting comparison is to the results of Ref. \cite{asthana25}.  They calibrated the $\dot{n}^{\rm esc}_{\rm ion}(z)$ in radiative transfer simulations of reionization to observed Ly$\alpha$ forest transmission values at $5\lesssim z \lesssim 6$ (see \S \ref{sec:intro}). They then used recent JWST measurements of $\xi_{\rm ion}$ to infer $f_{\rm esc}(z)$. They found average escape fractions of $\approx 4\%$ and $10\%$ at $z=5$ and 8, respectively, remarkably close to our values of 3\% and 12\% at those redshifts.   

Lastly, one of our key conclusions is that reionization feedback may not be necessary to account for the declining $\dot{n}^{\rm esc}_{\rm ion}$, depending on the still-uncertain slope of the decline.  This differs from the conclusion of Ref. \cite{2021MNRAS.507.6108O}, who attribute the decline entirely to reionization feedback in low-mass halos with dark matter masses $M_{\rm DM}\lesssim 2\times 10^9$ M$_\odot$.  We suggest that our different conclusions may owe to a combination of numerical effects: (1) Their star formation prescription allows cells with gas densities of 50 times the cosmic mean and temperatures below $20,000$ K to form stars.  This temperature threshold is similar to the peak temperatures behind ionization fronts (I-fronts) during reionization \cite{daloisio18}.  Hence, star formation in under-resolved regions of the simulation, where self-shielding is not well captured, may be artificially shut off immediately after I-fronts cross them, potentially exaggerating the effect of reionization feedback; (2) Ref. \cite{2021MNRAS.507.6108O} model the $f_{\rm esc}$ of each galaxy by adopting a fixed value for star particles (unity or 0.45, depending on the model) and ray tracing LyC radiation to the halo virial radius. In their simulations, escape fractions approach unity for the low-mass halos that are susceptible to reionization feedback. However, their grid resolution of 11.53 ckpc is too coarse to resolve the ISM structure that sets $f_{\rm esc}$.  Their large $f_{\rm esc}$ for low-mass halos amplifies the effects of reionization feedback on the emissivity.     

\begin{table*}[t]
    \centering
    \
    \label{tab:discussion_models}
    \small
    \begin{tabularx}{\linewidth}{@{}l c >{\raggedright\arraybackslash}X@{}}
        \hline
        Reference & Effective $f_{\rm esc}(z)$ & Notes \\ 
        \hline
        \noalign{\vskip 3pt}
        Yung et al. (2020)~\cite{2020MNRAS.tmp.1918Y}
        & $\displaystyle
            \frac{0.2985}
                 {1+\left(\frac{0.2985}{0.0381}-1\right)e^{-0.523(z-4)}}$
        & SAM model, used MCMC to jointly constrain $f_{\rm esc}$ and stellar feedback strength with Ly$\alpha$ forest $\dot{n}_{\rm ion}^{\rm esc}$ measurements of \cite{becker13} and \textit{Planck} 2016 $\tau_{\rm es}$ \cite{2016arXiv160503507P}, median of marginalized posteriors used here \\
    \noalign{\vskip 3pt}
    Haardt \& Madau (2012)~\cite{haardt12}
& $\min[1,\;1.8\times10^{-4}(1+z)^{3.4}]$ & Ionizing background model, matches Ly$\alpha$ forest $\Gamma_{\rm HI}$ at $z\simeq 2$--5 \cite{bolton07,becker06,faucher08}, increases with $z$ for reionization history consistent with Seven-Year \textit{WMAP} $\tau_{\rm es}$ \cite{jarosik11} \\
\noalign{\vskip 3pt}
Sharma et al.\ (2017)~\cite{sharma17}
& $0.045\,\left(\frac{1+z}{4}\right)^{1.1}$
&  Model built on \textsc{EAGLE} galaxies, observationally motivated $\Sigma_{\rm SFR}$--threshold for LyC leakage, luminosity-weighted average $f_{\rm esc}$ fit with power-law in $(1+z)$ \\
\noalign{\vskip 3pt}
    
        \hline
    \end{tabularx}
    \caption{Summary of the parametric models for $f_{\rm esc}(z)$ shown in the right panel of Figure \ref{fig:comparisons}.}
\end{table*}

\section{Conclusion}
\label{sec:conclusion}

Measurements of the Ly$\alpha$ forest mean transmission provide relatively stringent constraints on the end of reionization.  Independent attempts to model these measurements suggest that the ionizing emissivity started to decline during the latter half of reionization.  In this paper, we have used the Santa Cruz SAM framework to consider two proposed physical mechanisms that could have driven the decline: (1) the suppression of star formation in low-mass halos by photoheating of the IGM (``reionization feedback'') and (2) a declining escape fraction associated with the diminishing effectiveness of stellar feedback in facilitating LyC leakage.   Our main findings are summarized as follows:

\begin{itemize}
\item 
In the model of \textsc{Noh \& McQuinn} considered here, reionization feedback suppresses star formation primarily in low-mass halos with $M_{\rm halo}\lesssim 10^9$ M$_\odot$.  This flattens the slope of the rest-frame UV luminosity function at magnitudes fainter than $M_{\rm UV} \simeq -13$, near the faintest luminosities probed by recent JWST measurements exploiting strong lensing magnification in galaxy clusters \cite{2026OJAp....966816A}. 

\item 
In a scenario where escape fractions at fixed $M_{\rm halo}$ do not evolve considerably over $5 < z < 8$,  producing a declining emissivity requires that the population-averaged escape fraction is close to zero for halos with $M_{\rm halo}\gtrsim 10^{10}$ M$_\odot$ because of the steeply rising ionizing photon production rates of these galaxies. Under these conditions, reionization feedback may contribute to the downward evolution. However, in this scenario, potentially all of the decline can be driven by a combination of hierarchical assembly and stellar feedback shifting the ionizing photon production toward larger halo masses.  
\item 
Based on the results of recent radiation hydrodynamics simulations, the population-averaged escape fraction is expected to evolve rapidly at the epochs of interest here ($z\sim 6$) because of the decreasing effectiveness of star formation and its feedback in facilitating LyC leakage \cite[e.g.][]{2020MNRAS.498.2001M, 2022MNRAS.515.2386R, 2026arXiv260614671S}.  We have implemented into our SAM framework a prescription for modeling this evolution from the \textsc{SPHINX} simulation project \cite{2022MNRAS.515.2386R}. After an overall rescaling by a constant factor, we find that the predicted $f_{\rm esc}$ evolution matches the required redshift evolution of the emissivity remarkably well. The main results of this paper are shown in Figure \ref{fig:nion_sphinx}.  

\item 
In a model with both \textsc{Noh \& McQuinn} reionization feedback and \textsc{SPHINX} $f_{\rm esc}$ evolution, we find that reionization feedback contributes only mildly to the emissivity evolution. In fact, within the (still large) uncertainties in the shape of the emissivity's redshift-dependence, the decline may have been driven entirely by $f_{\rm esc}$ evolution.  In this case, the timing of the decline with the last phases of reionization is coincidental.   

\item 
The evolution of the Ly$\alpha$ forest mean transmission alone cannot be used to distinguish between the models considered here.

\end{itemize}

Although the presence of a flattened or declining emissivity near the end of reionization appears to be qualitatively robust to modeling uncertainties \cite{2024MNRAS.531.1951C}, the {\it rate of decline} is sensitive to uncertainties in both the modeling and observational transmission measurements. For instance, some models require a steep decline while in others the evolution is flatter (see Fig. \ref{fig1}). Some of this variation owes to uncertainties in the first half of the reionization history.  Generally, a more compressed timeline (shorter duration) for reionization will require a more rapid evolution in the emissivity to maintain consistency with the forest measurements. As pointed out in \cite{2024MNRAS.531.1951C}, additional modeling uncertainties include assumptions about the spectrum of the ionizing background and the distribution of \textsc{H~i} columns around star forming galaxies. The conclusions of the current paper, however, are aimed at a broader level than these details. Many reionization modelers have considered the non-monotonic evolution of the emissivity in their models to be an un-physical artifact of the simulation methods.   As was originally recognized by \cite{2022MNRAS.515.2386R}, our work supports the view that a turnover in the emissivity is actually expected from the physics of $f_{\rm esc}$.

\acknowledgments
The authors thank Matt McQuinn, George Becker, and Brian Siana for helpful discussions and comments on this manuscript. AD used ChatGPT GPT-5.6 Sol to aid the coding of the \textsc{SPHINX} $f_{\rm esc}$ model and plotting scripts, and to speed up elements of post-processing codes. AD checked these codes and takes full responsibility for them.  All scientific analysis/interpretation, manuscript text, results, and conclusions are the authors' own.  AD acknowledges support from grant NSF AST-2045600. LYAY acknowledges support from the Taiwanese Ministry of Education (MoE) Yushan Fellowship (MOE-115-YSFMS-0002-005-P1).  
CC acknowledges support from the Kavli Institute of Cosmology at Cambridge and the Beus Center for Cosmic Foundations at Arizona State University.  The Flatiron Institute is supported by the Simons Foundation. 
Computations were performed on Expanse at the San Diego Supercomputer Center (SDSC) through allocations PHY240332 and PHY210041 from the NSF (ACCESS) program.

\appendix

\section{On the Correlation Between Reionization Redshift and Halo Mass}
\label{app:zre}

Our implementation of the \textsc{Noh \& McQuinn} reionization feedback model makes the approximation that the probability distribution for $z_{\rm re}$ is independent of halo mass.  This approximation neglects the effect of halo clustering on $z_{\rm re}$.  The most massive halos formed in over-dense environments, surrounded by other massive halos, and were among the first to be reionized in modern models of reionization.  Here we check the accuracy of our approximation. 

For this purpose, we use the \textsc{Late Start/Late End} radiative transfer simulation of reionization from Ref. \cite{2024arXiv240902989C}. This simulation was run with the \textsc{FlexRT} adaptive ray tracing code \cite{2024arXiv240904521C} in a cosmological volume with box size $L_{\rm box}=200~h^{-1}$cMpc. We refer the reader to Ref. \cite{2024arXiv240902989C} for more technical details about the simulation. We extracted the $z_{\rm re}$ for each halo in the simulation using values stored on the uniform $N_{\rm RT} = 200^3$ radiative transfer grid. 

\begin{figure}
    \centering
    \includegraphics[width=0.8\columnwidth]{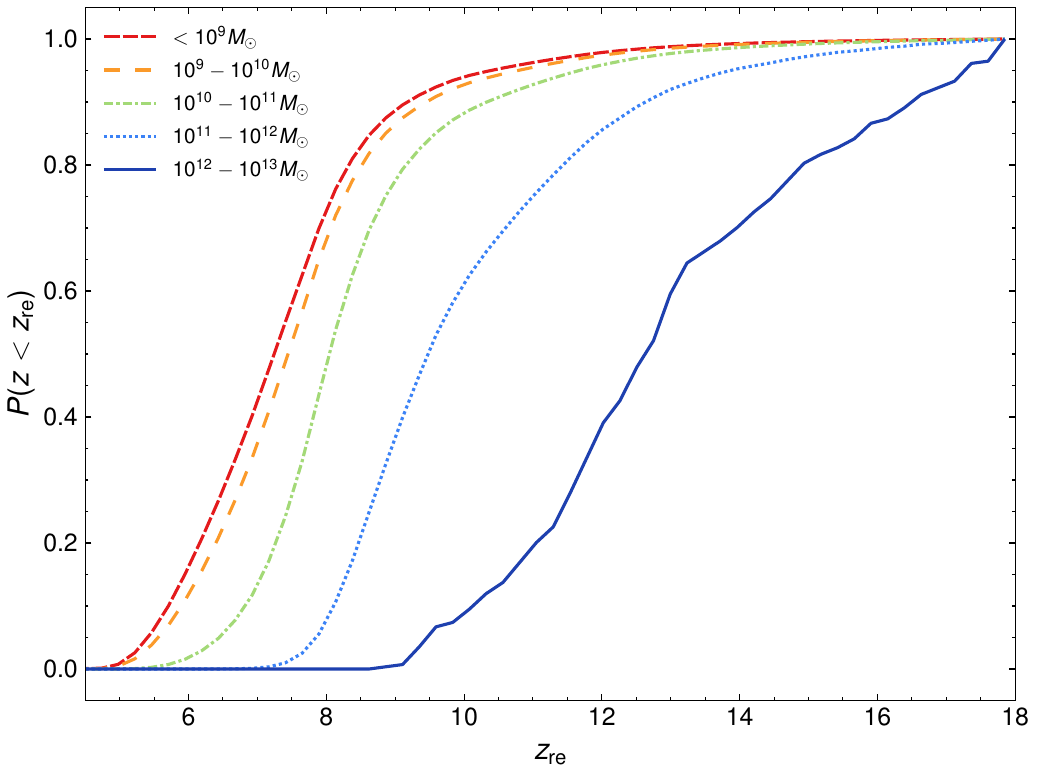}
    \caption{Cumulative probability distribution for the reionization redshift of a halo, $z_{\rm re}$, from the \textsc{Late Start/Late End} radiative transfer simulation of \cite{2024arXiv240902989C}, binned by halo mass.  At low masses ($M_{\rm halo}\lesssim 10^{10}$ M$_\odot$), the distribution is only modestly dependent on halo mass. These results support the approximation made in our implementation of the \textsc{Noh \& McQuinn} model, where we draw random values for $z_{\rm re}$ from a distribution that is independent of halo mass. In that model, only halos with masses below $M_{\rm halo}\sim 10^9$ $M_\odot$ are affected by reionization feedback, safely in the regime where we may negelect the mass dependence of the $z_{\rm re}$ distribution. }
    \label{fig:Reo_timesappendix}

\end{figure}

Figure \ref{fig:Reo_timesappendix} shows the cumulative probability distribution of $z_{\rm re}$ for different halo mass bins, as indicated in the plot legend. For the two lowest mass-bins, i.e. M$_{\rm halo}$ $< 10^{10}$ M$_\odot$, the cumulative distributions are similar, indicating that the $z_{\rm re}$ distribution is only weakly dependent on halo mass below this mass threshold. On the other hand, the $z_{\rm re}$ distribution has a very strong dependence on halo mass above this threshold, with the most massive/highly biased halos exclusively reionized during the earliest phases of reionization.  What matters for our purposes, however, is that the $z_{\rm re}$ distribution for the low-mass halos affected by reionization feedback is only weakly dependent on halo mass.  As shown in Figure \ref{fig:Mchar}, the threshold below which halos are subject to reionization feedback is $\sim 10^{9}$ M$_{\odot}$ in the \textsc{Noh \& McQuinn} model, safely in the regime where the $z_{\rm re}$ distribution of Figure \ref{fig:Reo_timesappendix} shows little mass dependence. These results lead us to conclude that neglecting the halo mass dependence is a reasonable approximation for our purposes.

\section{Disentangling the Effects of Cosmological Evolution from Stellar and AGN Feedback}
\label{app:feedback}

\begin{figure}
    \centering
    \includegraphics[width=\columnwidth]{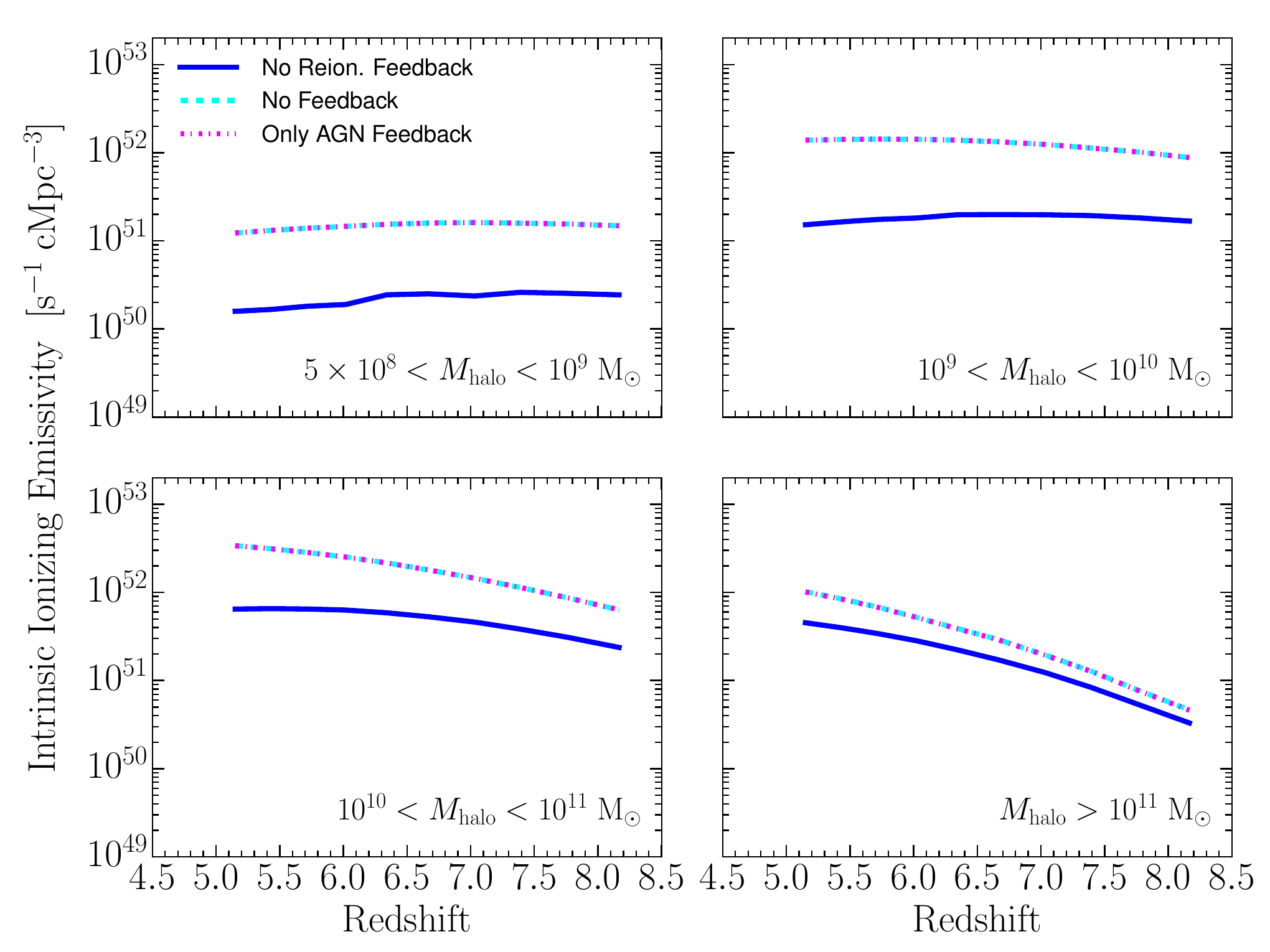}
    \caption{Testing the effects of turning off stellar and AGN feedback. As in Figure \ref{fig:nion_int}, the panels show intrinsic ionizing emissivities in different halo mass bins.  In the \textsc{No Feedback} model, we turn off stellar and AGN feedback (in addition to reionization feedback).  In the \textsc{Only AGN Feedback} model, stellar and reionization feedback are both off, but AGN feedback is left on. AGN feedback in the Santa Cruz SAM has an insignificant effect on the emissivity at these redshifts.  Although stellar feedback has a strong effect on the normalization of the emissivity, the slopes are set largely by cosmological evolution, i.e. halo growth/mergers.  }
    \label{fig:nion_feedbackappendix}
\end{figure}

This appendix addresses two main questions: (1) Figure \ref{fig:nion_int} shows that the intrinsic ionizing emissivities in the lowest two mass bins exhibit a flat or mildly declining evolution, even in the \textsc{No Reion. Feedback} model.  What are the roles of stellar feedback and cosmological evolution (accretion rates and halo abundances) in setting this evolution?; (2) What is the role of AGN feedback at these redshifts?

\begin{figure}
    \centering
    \includegraphics[width=\columnwidth]{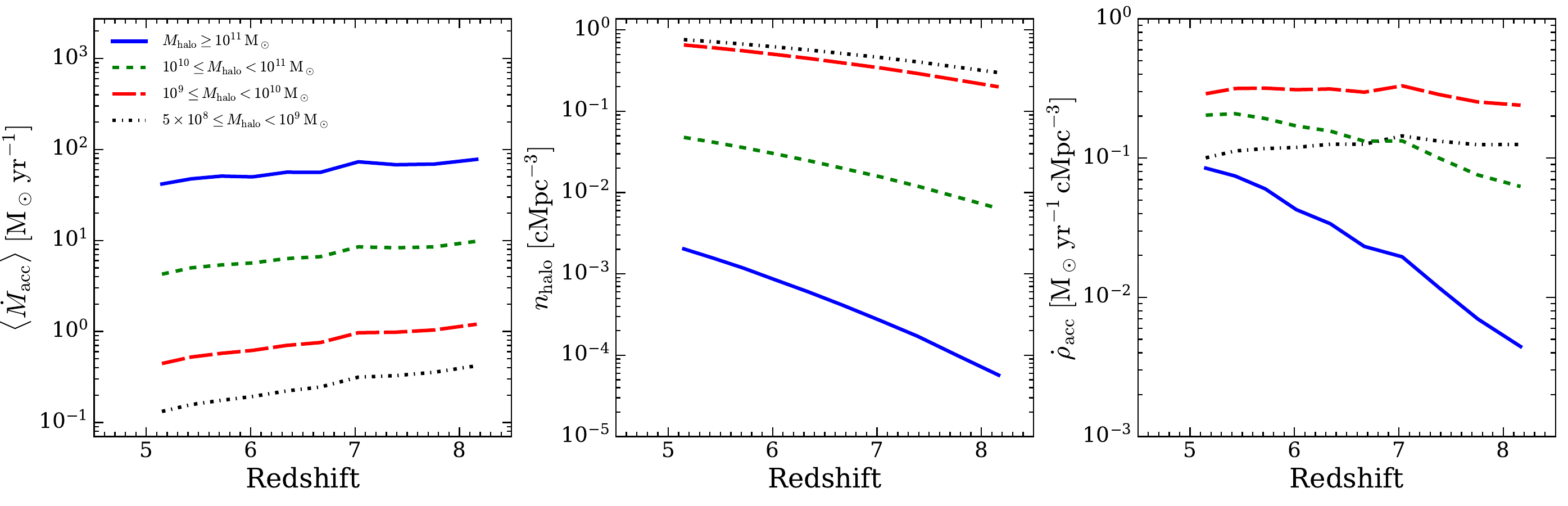 }
    \caption{The effects of halo growth and hierarchical assembly on the evolution of the intrinsic emissivity in the \textsc{No Reion. Feedback} model.  {\bf Left:} The average accretion rate of pristine gas, binned by halo mass, for central galaxies in the SAM. Here, the pristine gas accretion rate is defined to be the smooth (excluding resolved progenitors) halo mass growth rate times the cosmic baryon fraction. {\bf Middle:}  The evolution of the halo number density for central galaxies.  {\bf Right:}  The pristine gas accretion rate per unit volume, $\dot{\rho}_{\rm acc}$.  Whether $\dot{\rho}_{\rm acc}$ grows with cosmic time depends on whether the halo density grows sufficiently fast to counteract the declining average accretion rate. In the most massive halo bin, the halo abundance grows rapidly enough to drive a steep growth in $\dot{\rho}_{\rm acc}$. In the least massive bin, the halo abundance growth is insufficient.   These results suggest that cosmological evolution in the supply of fresh gas to halos plays an important role in determining whether the intrinsic emissivity is rising or flattening/declining in a given mass bin.   }
    \label{fig:cosmo_evolution}
\end{figure}

To address these questions, we have performed two additional test runs.  In one run, which we label \textsc{No Feedback}, we turn off stellar and AGN feedback.  In the second run, we turn on only AGN feedback.  Figure \ref{fig:nion_feedbackappendix} shows the intrinsic ionizing emissivities in these runs, compared against the \textsc{No Reion. Feedback} run from the main text. First, we see that the AGN feedback prescription in the SAM has a negligible effect on the emissivity at these redshifts. In all mass bins, the overall emissivity is larger in the case where stellar feedback is shut off.  The slope of the redshift evolution in the lowest mass bin is similar to that of the \textsc{No Reion. Feedback} run, indicating that the slope there is set primarily by cosmological halo growth, rather than by stellar feedback.  Turning off stellar feedback does have a steepening effect on the emissivities in the other mass bins, however. In the $10^9 < M_{\rm halo} < 10^{10}$ M$_\odot$ bin, the emissivity even changes from mildy declining to mildy rising.  

Now we examine the role of halo accretion rates and abundances in shaping the emissivity evolution within the \textsc{No Reion. Feedback} model.  The left panel of Figure \ref{fig:cosmo_evolution} shows the mean pristine gas accretion rate for central galaxies in the same 4 halo mass bins as in Figure \ref{fig:nion_int}. Specifically, we define the pristine gas accretion rate to be the ``smooth'' (excluding resolved progenitors) halo mass growth rate times the cosmic baryon fraction.  Within the SAM framework, this is the infall rate of unenriched hot gas in a halo.  The left panel shows that this accretion rate declines on average in each mass bin. What matters for the emissivity, however, is the evolution of the accretion rate per unit volume.  A key question, then, is whether halo number density in each bin grows steeply enough with time to counteract the declining accretion rates.  The middle panel of Figure \ref{fig:cosmo_evolution} shows the evolution of the halo number density in each mass bin.  Finally, the right panel shows the pristine gas accretion rate density.  In the most massive bin, the halo density is growing rapidly enough to drive a steep increase in the accretion rate density.  This is responsible for producing a steeply rising emissivity in that bin.  By contrast, for the lowest mass bin, the halo density does not grow steeply enough to counter the declining accretion rates (on average).  Hence the accretion rate density declines in that mass bin, driving a decline in the emissivity.   

In summary, these results suggest that the intrinsic emissivity's slope is shaped by a combination of halo accretion rates, halo abundance, and stellar feedback.          
\bibliographystyle{JHEP}
\bibliography{References}
\end{document}